\documentclass[11pt]{article}

\usepackage[latin1]{inputenc}
\usepackage{amsthm,amsmath,amsfonts} 
\usepackage{amssymb}
\usepackage{mathrsfs}

\usepackage{eurosym}
\usepackage{verbatim} 
\usepackage[english]{babel}
\usepackage[T1]{fontenc}
\usepackage{bbm,bm}

\usepackage{lmodern}
\usepackage{blindtext}

\usepackage{tabularx}
\usepackage[table,xcdraw]{xcolor}
\usepackage{float}

\usepackage{graphicx, floatflt} 
\usepackage[labelformat=simple]{subcaption}

\DeclareCaptionLabelFormat{subcaptionlabel}{\normalfont(\textbf{#2}\normalfont)}
\usepackage{color}
\usepackage[export]{adjustbox}

\usepackage{enumitem}

\usepackage{multirow}
\usepackage{booktabs}
\usepackage{adjustbox}
\usepackage{blindtext}
\usepackage{esint}
\usepackage[title]{appendix}

\theoremstyle{definition}

\graphicspath{{Figures/}}

\usepackage{authblk}

\title{\bf Enhancing CVaR portfolio optimisation performance with GAM factor models }
\author[1,*]{Davide Lauria}
\author[2]{W. Brent Lindquist}
\author[2]{Svetlozar T. Rachev}

\affil[1]{\small Department of Economics, Statistics and Finance, University of Calabria, Arcavacata di Rende, Italy.
}

\affil[2]{\small Department of Mathematics \& Statistics, Texas Tech University, Lubbock, TX, U.S.A.
brent.lindquist@ttu.edu; zari.rachev@ttu.edu}

\affil[*]{Corresponding author, davide.lauria@unical.it}

\begin{document}
\maketitle

\noindent\textbf {Abstract}
We propose a discrete-time econometric model  that combines autoregressive filters with factor regressions to predict stock returns for portfolio optimisation purposes. 
In particular, we test both robust linear regressions and general additive models on two different investment universes composed of the Dow Jones Industrial Average and the Standard \& Poor's 500 indexes, and we compare the out-of-sample performances of mean-CVaR optimal portfolios over a horizon of six years. The results show a substantial improvement in portfolio performances when the factor model is estimated with general additive models.\\


\section{Introduction}\label{sec:intro}                                             


Modern portfolio theory provides quantitative techniques that can be used to form a portfolio of financial securities that is optimal in terms of a particular set of objectives. 
Since the choice of the portfolio depends on the expectations of the uncertain realisation of future financial returns, the value of the portfolio is described by a stochastic process, and the problem is formulated as a stochastic optimisation; this optimisation is performed  on the basis of some risk--reward trade-off. 
From a practical perspective, portfolio optimisation requires the definition of i) a stochastic model for the evolution of the returns, ii) an objective function that describes the investor's preferences concerning the risk--reward trade-off, and iii) a set of constraints that represent real market features such as transaction costs and limitations on short positions.
In the seminal mean--variance (MV) approach of Markowitz \cite{Markowitz_1952,Markowitz_1956}, for instance,  the investor seeks the minimal variance of the future portfolio distribution for any given level of its expected return. In this case, financial returns are supposed to be governed by a Gaussian probability law, and  no restrictions on the set of possible strategies are imposed; under such assumptions, Markowitz's problem can be written as a quadratic mathematical programming problem and analytical solutions can be found.
Konno and Yamazaki \cite{Konno_1991} proposed the use of the mean-absolute deviation to take advantage of a linear programming reformulation and solve large-scale portfolio problems with dense covariance matrices. This approach was later generalised  by Ogryczak using standard semi-deviations \cite{Ogryczak_1999} and a multiple criteria  model \cite{Ogryczak_2000}. 

From a methodological point of view, an axiomatic definition of a risk measure has been proposed by Artzner at al. \cite{Artzner_1999}. They provide basic properties that such operators should satisfy in order to represent rational preferences regarding risk levels; functionals that satisfy the axioms are called `coherent' (see also \cite{Delbaen_2002, Follmer_2010}). The asymmetric nature of the distribution of returns has motivated the development of a new class of risk measures that focus on extreme losses, such as the expected shortfall \cite{Acerbi_2002} and the conditional value-at-risk \cite{Rockafellar_2000, Pflug_2000}. 
Among the coherent risk measures introduced in the literature, the conditional value-at-risk (CVaR) has interesting features that make it a flexible choice for the risk measure.
The CVaR is indeed consistent with the second-degree stochastic dominance relation \cite{Ogryczak_2002}, and it can be efficiently linearised  \cite{Rockafellar_2000} on discrete-state probability spaces; it has also been empirically proven to improve the performances of MV portfolios in practical financial problems \cite{Andersson_2001, Mansini_2003, Topaloglou_2002, Stoyanov_2013}.\\

After the choice of the risk measure, the out-of-sample performance of the optimal policy depends on both the definition of the stochastic model that represents the evolution of the financial returns and the constraints that determine the set of feasible investment  policies.
The Gaussian assumption in Markowitz's MV has been proven to be unrealistic \cite{Fama_1963, Mandelbrot_1963}, and it is now well known that financial returns, at least for frequencies that are less than quarterly, exhibit serial dependence in the volatility (volatility cluster); their probability distribution has tails that are heavier than normal, and are asymmetric (see \cite{ Rachev_2003} and the references therein). It has also been proven that the impact of transaction costs \cite{Magill_1976, Liu_2004, Lobo_2007} and limits on short positions  \cite{Brennan_2010, Grullon_2015} substantially affect the realised returns of the optimal strategies. 

A powerful method for solving the stochastic optimisation problem involves performing the time and state discretisation of the stochastic process.
A discrete state representation of the uncertainty allows writing the stochastic portfolio problem as a large deterministic optimisation (see \cite{Dupavcova_1999, Ruszczynski_2009}).
Furthermore, in many cases, and under mild hypotheses, such a deterministic representation can be made convex or linear, guaranteeing the uniqueness of the solution. 
Usually, the uncertainty of future financial returns is represented either directly with empirical data in non-parametric methods (data-driven optimisation) \cite{Rockafellar_2000,Mansini_2007} or by defining a parametric model \cite{Topaloglou_2002,Sahamkhadam_2018}.
In both cases, the model is usually estimated based on the information provided by historical financial returns without considering exogenous factors such as economic indicators or financial variables of the companies whose stocks are in the asset universe.\\

Another group of studies has focused on explaining market excess returns on the basis of exogenous variables called `factors'. 
The first multi-factor model was proposed by Rosenberg \cite{Rosenberg_1974}, who noticed that in general, portfolio managers and financial analysts consider the economic and financial characteristics of the company to have an impact on asset returns. 
Ross \cite{Ross_1976} regressed the return of an asset on a set of factors in order to relax the conditions behind the Capital Arbitrage Pricing Model (CAPM) of Sharpe, Linter, and Mossin \cite{Sharpe_1964,Lintner_1965,Mossin_1966} through no-arbitrage arguments. 
The theoretical grounds of Ross' work did not receive much attention from an operative perspective until Fama and French  \cite{Fama_1992} found a statistical relation between factors that describes the financial end economic variables of the security, such as the capitalisation and the book-to-market ratio. 
Later on, a substantial group of studies proposed similar models to explain excess returns, adding more factors and testing the models on different markets (see, for instance, \cite{Carhart_1997,Titman_1997,Fama_2015,Hou_2015,Markowitz_2021}).
Factor models are now increasingly used in the financial industry for either risk management or active portfolio management, and they are often referred to as BARRA-type factor models, after the company BARRA, Inc., which provided a technical illustration of this method \cite{Sheikh_1996}. 
A recent econometric approach to estimating factor loadings in high-frequency markets was proposed by Dai at al. \cite{Dai_2019}.\\

Jegadeesh and Titman \cite{Jegadeesh_1993} found that equity markets exhibit medium-term return momentum, which means that securities that have performed the best (worst) over the last three-to-twelve months have a tendency to show the same performance during the next period of the same length. 
They showed that portfolio strategies that take advantage of such features show abnormal excess returns. This analysis has been further tested \cite{Jegadeesh_2001,Rouwenhorst_1998,Griffin_2005,Hanauer_2023}, confirming the potential profitability of such strategies.
This approach shares methodological features with the so-called \textit{technical analysis}.
Technical analysis can be defined as a set of methods, both graphical and numerical, that extrapolate the future directions of stock prices from previous data; these methods mostly predict the prices themselves and volumes \cite{Schwager_1995}.
The lack of a clear methodology and the subjective nature of many of the tools in the technical analysis framework have created a sort of mistrust among academics despite the extensive use of such techniques among financial professionals.
However, some technical indicators/techniques have been tested using statistical tests, and the majority of such studies claimed to have found positive results (see, for instance, \cite{Lo_2000,Park_2007} and the references therein).
Positive findings have encouraged the development of a new class of models that incorporate traditional technical tools and machine learning and statistical approaches (see, for instance, \cite{Wei_2011,Ticknor_2013,Lin_2018,Nazario_2017} for a literature review).\\

In this work, we propose a model that incorporates factors belonging to the fundamental, momentum, and technical analysis classes into a traditional time-series approach in order to simulate future returns; the objective is to enhance the out-of-sample performance of a mean-CVaR portfolio optimisation run over the sampled scenarios.
We first estimate an autoregressive filter for both the conditional mean and variance on the series of daily returns for each asset in the investment universe. 
 A factor regression is then added to provide auxiliary information in addition to the innovations obtained from the time-series model. 
The dependence between securities is estimated by fitting a suitable multivariate probability distribution belonging to the generalised hyperbolic (GH) class to the collection of standardised residuals from each individual autoregressive filter. 
In particular, we tested two different factor models: a robust linear regression (RLR) and a general additive model (GAM). 
RLR was introduced in order to limit the impact of outliers on the estimation of the parameters of a linear regression \cite{Tukey_1960,Huber_1964,Huber_2011}.
GAMs were developed by Hastie in order to provide more flexibility to the possible functional relationship between variables than the linear regression model.

The linear function is replaced by a sum of smooth functions which need to be estimated using an iterative procedure \cite{Hastie_1987}. GAMs have been used in many applications, but they have not, to the best of our knowledge, been used as factor models for asset returns, especially for portfolio optimisation purposes.\\

The purpose of this paper is twofold. The first objective is to test if factor models can enhance the out-of-sample performance of time-series models for numerical portfolio optimisation; the second is to use GAMs as factor models in finance. The two purposes should be framed in the context of strategic portfolio optimisation: the goal is not to find a stable relationship between factors and financial returns over time, but to extract information in order to enhance the stability and the profitability of the optimal strategy.
The out-of-sample analysis has been performed on the asset universe of the Dow Jones Industrial  Average index and on a subspace of the Standard \& Poor's 500 index. This choice is motivated by the fact that these markets are highly efficient, and positive results indicate that this method should be extended to more inefficient markets to obtain improved performances. 
%
The rest of the paper is organised as follows. In Section \ref{sec:Model}, we describe the alternative models that will be tested. In Section \ref{sec:EmpiricalAnalysis}, the out-of-sample results from the case study are presented.


\section{The Model} \label{sec:Model}                                            

In this section, we present the models used to forecast future returns and compute optimal portfolio performances. We consider the case in which the investor starts at date $t_{0}$ with $X_{0}$  dollars to be completely invested  in $I$ securities, indexed by the set ${\cal I}=\left\{ 1, \dots, I\right\}$ and traded in some financial market along a horizon of $T$ days. At each date $t$, the investor knows the matrix $R_{t}$ of size $T_{w} \times I$ that contains the previous $T_{w}$ compounded returns of each asset; in other words, the parameter $T_{w}$ defines the time window in which we estimate the model. 
We also assume that we have at time $t$ a sequence of matrices $F_{i,t}$, for $i=1,\dots,I$, which each have size $T_{w} \times K$, which store the values of the $K$ factors (exogenous variables) during the time interval $\left[ t - T_{w} , t-1\right]$. In this way, the factor model at $t$ will be estimated using the previous $T_{w}$ data without considering the current value; this will allow us to forecast the future one-step-ahead returns without forecasting the factors.   
The agent will decide, at each date $t$, with $t \in \left\{ t_{0}, t_{0}+1,\dots, t_{0}+T-1 \right\}$, the proportion $\theta_{i,t}$ to be invested in the $i$-th asset, with $i \in 1,\dots,I$. In this work, we do not allow short-selling positions, so the portfolio proportions must be positive and add up to one: $\theta_{i} \ge 0 , \forall i \in {\cal I }$, and $\sum_{i=1}^{I}\theta_{i} = 1$.
The model estimation process and the resulting optimal decision at each date $t$ are described by  the following set of actions:

\begin{enumerate}

\item We first estimate an ARMA-GARCH filter for each column of the matrix $R_{t}$ of historical compounded returns (see \cite{Hamilton_2020} for a detailed explanation) in order to obtain an endogenous model for each of the $I$ assets.
In particular, by introducing the lag operator $L$ and the error process $\left\{\epsilon_{i,t} \right\}$ with zero mean and variance $\sigma_{i,t}^{2}$, the time-series model  for the $i$-th asset's log return process  $\left\{ r_{i,t} \right\} $ can be formally defined by the following two processes:
\begin{align} 
     & \left( 1 - \alpha_{i,1} L -\alpha_{i,2} L^{2} \right) r_{i,t} =  c_{i,1} + \left( 1 + \beta_{i,1}L + \beta_{i,2} L^{2} \right) \epsilon_{i,t}, \label{eq:sarima} \\
     & \left( 1 - \phi_{i,1} L - \phi_{i,2} L^{2} \right)\sigma_{i, t}^{2} = c_{i,2} + \left( 1 + \gamma_{i,1}L + \gamma_{i,2} L^{2} \right) \epsilon_{i,t}^{2} . \label{eq:garch}
\end{align}
For each asset in ${\cal I }$, we estimate all the combinations of parameters different from zero and select the best model based on the BIC information criterion (see \cite{Schwarz_1978}).
The output is a vector of innovations $h_{i,t} := \frac{ \epsilon_{i,t} }{\sigma_{i, t}}$ of size $T_{w}$ that is obtained from the $i$-th filter.

\item The vector $h_{i,t}$ is then passed through a regression described by a function $g\left( \cdot \right)$ of the type 
          \begin{align}
                  & h_{i,t} = g\left( F_{i,t} \right) + \xi_{i,t},
          \end{align}
          where $\xi_{i,t}$ is a random variable with zero mean and unit variance. In this work, we consider two different functionals for $g\left( \cdot \right)$, which will be described in 
          subsection \ref{subsec:FactorModels}.

\item The $I$ vectors of regression residuals $\xi_{i,t}$ are then collected in the size $T_{w} \times I$ matrix $\Xi_{t}$. 
      At this point, we are able to estimate a multidimensional \textit{normal inverse Gaussian} (NIG) distribution on the set of historical residuals $\Xi_{t}$. The choice of the NIG distribution is motivated by its flexibility in matching the empirical properties of financial returns, as shown in \cite{BN97,Biglova2014}.  A brief description of the NIG distribution and the EM algorithm to estimate it can be found in Appendix \ref{sec:App_B}. 

\item We are now able to simulate $S$ random numbers from the estimated multivariate NIG distribution, and thus we can produce 
         a matrix of simulated residuals $\hat{ \Xi }_{t+1}$ in order to obtain the one-step-ahead simulated innovations 
         $\hat{ h }_{i,t+1} $, for $i = 1, \dots, I$. By  passing these innovations through the filters described by equations
          (\ref{eq:sarima})--(\ref{eq:garch}), we can finally obtain the $S \cdot I$ next-period simulated returns 
          $\hat{ r }^{ (s) }_{ i , t+1}$, for each $i \in I$ and $s=1, \dots S$.

\item We consider next the optimal portfolio problem, which is defined as a convex combination of the expected reward and  the risk functional  
         CVaR${}_\beta$ introduced by Rockafellar and Uryasev \cite{Rockafellar_2000}. 
         The  optimal portfolio problem at date $t$ can be formulated as the following mathematical programming problem:
          \begin{equation}
                      \min_{\theta,\nu } \left\{ -\alpha \theta' \mu+  ( 1- \alpha) \left[ \nu +S^{-1} \left( 1 - \beta \right)^{-1} \sum_{s=1}^S \left[ -x_s- \nu \right]^{+} \right] \right\},
                      \label{eq:MCVeR_objf}
          \end{equation}
          where
          \begin{equation}
                    x_s =  \sum_{i=1}^I \hat{ r }^{(s)}_{ i, t+1}  \theta_i, \text{ for } s =1,\dots,S,
          \end{equation}
          subject to
          \begin{align}
                    \theta_{i} &\ge 0, \text{ for } i =1,\dots,I,\\ 
                    \sum_{i=1}^{I} \theta_{i} &= 1,\\
                    \nu &\in \mathbb{R} \label{eq:MCVaR_const_4},
          \end{align}
          for a particular choice of $\alpha  \in [0,1]$.
          We use 0.99 as the value of the significance coefficient $\beta$.
          Equations (\ref{eq:MCVeR_objf})--(\ref{eq:MCVaR_const_4}) can be reduced to a linear problem, as described by Rockafellar and Uryasev \cite{Rockafellar_2000}.

\end{enumerate}

\subsection{Factor Models} \label{subsec:FactorModels}    

We consider two  alternative specifications for the functional form of  $g\left( \cdot \right)$, which explains the relation between the innovations of the autoregression filter defined by equations (\ref{eq:sarima})--(\ref{eq:garch}) and a set of $K$ factors.
The first factor model is based on \textit{robust linear regression} (RLR) (see, for instance, \cite{Li_1985}).
The motivation behind RLR is to reduce the bias in ordinary least squares when the distribution of the residuals shows heavy tails and the outliers are not negligible. 
The idea is to weight the observations: by reducing the importance of outliers, this type of regression is able to improve the efficiency compared to the case in which the outliers are simply removed.
There are several approaches to implementing robust estimation; here, we focus on the \text{M-estimation} method and use Huber and Tukey's bi-weight objective functions\footnote{In the case study, we used the  function \textit{rlm} from the R package `MASS' \cite{Venables_2002}.}.
A detailed description of the robust regression applied in this work can be found in Appendix \ref{sec:App_C}.

As the second test, we applied the generalised additive model (GAM) introduced in \cite{Hastie_1987}.
In general, the GAM expresses the expected value of the response variable as the sum of smooth functions of the dependent variables. The advantage of such a class of models is their flexibility in capturing non-linear relations without imposing pre-established parametric functional forms. Of course, one should still find a suitable representation of the smooth functions and, furthermore, define the parameters that control the smoothness\footnote{
There are, however, techniques that allow one to obtain the best smoothness according to the data; see, for instance, Ch. 3 in \cite{Wood_2011}.
}. 
In this work, we check the simplest case, where each smoothed  function is a P-spline that takes as input one single factor\footnote{This implementation has been carried out using the \textit{gam} function in the R package `mgcv' (see \cite{Wood_2011} for a detailed explanation).}. The model is further explained in Appendix \ref{sec:App_D}.

\subsection{Description of Factors} \label{subsec:FactorDescription}    
We describe here the 20 factors used in the regressions by dividing them into three main categories: momentum, fundamental, and technical indicators.
The first category includes factors that indicate the momentum of a single stock during a given recent period. In particular, we have used the Bloomberg Relative Strength Index (\textit{RSI})  calculated on four different time intervals: 3, 9, 14, and 30 days, respectively. The RSI is computed as follows:
\begin{align}   
	\textit{RSI} = 100 - \left[ 100 / \left( 1 + \frac{ \textit{Avg Up} }{ \textit{Avg Down} } \right) \right] ,
\end{align}
where \textit{Avg Up} is  the average of all day-on-day changes when the security closed up for the day during the interval and \textit{Avg Down} is the average of all down changes for the period. \textit{RSI}  is defined on a scale between 0 and 100.
We also considered, as a longer horizon momentum indicator, the Bloomberg Relative Share Price Momentum, which represents the percentage change over the last six months in the one-month moving average of the share price relative to a benchmark index (in this case, the DJIA index).

The second category contains four classical fundamental financial indicators: the price-to-earnings ratio (P/E), which is the ratio between the price of a stock and the company's earnings per share\footnote{The earnings per share (EPS)  measure is computed by dividing the income available to common stockholders by the weighted-average number of common shares that are outstanding during the period.}, the price-to-book ratio (P/B), the price-to-cash-flow ratio (P/C), which is the stock's price divided by the cash flow per share, and the price-to-sales ratio (P/S), which is defined as the ratio between the stock's last price and the sales per number of average outstanding shares.
In addition, for each of these four fundamental factors, we also considered the ratio between the actual value and the average over the past five years.
Another indicator is the EPS forecast, which has been shown to be informative of future price changes. In this work, we used the Bloomberg-calculated or company-reported EPS, which follows the Bloomberg's adjusted EPS consensus estimate.  
Finally, the last fundamental indicator is the \textit{Environmental, Social, and Governance} (ESG) score provided by the financial data provider Refinitiv. 
The ESG score is a weighted average of three indicators that represent the environmental and social impact of a company's activities and the ethical quality of its governance. Although the ESG score is not a traditional fundamental indicator, its importance in the overall valuation of a company is growing at a fast pace.  
\begin{table}[]
\centering
\scalebox{0.8}{
\begin{tabular}{clll} \hline \\ 
Index  & Factor Name& Source &Category\\ \hline
1 &RSI 3 days                                                     &  Bloomberg& Momentum     \\
2 &RSI 9 days                                                      &  Bloomberg& Momentum     \\
3 &RSI 14 days                                                    &  Bloomberg& Momentum     \\
4 &RSI 30 days                                                    &  Bloomberg& Momentum     \\
5 & Relative Share Price Momentum                     &  Bloomberg& Momentum    \\
6 &Price-to-earnings ratio                                &  Bloomberg& Fundamental    \\
7 &Price-to-book ratio                                      &  Bloomberg& Fundamental    \\
8 &Price-to-cash-flow ratio                                       &  Bloomberg& Fundamental    \\
9 &Price-to-sales ratio                                      &  Bloomberg& Fundamental   \\
10&Price-to-earnings ratio / 5-year average   &  Bloomberg&  Fundamental  \\
11&Price-to-book ratio  / 5-year average   &  Bloomberg&  Fundamental  \\
12&Price-to-cash-flow ratio / 5-year average   &  Bloomberg& Fundamental  \\
13&Price-to-sales ratio    / 5-year average   &  Bloomberg&  Fundamental   \\
14& EPS forecast       & Bloomberg&  Fundamental   \\
15&ESG total score from Refinitiv                        &Refinitiv      & Fundamental      \\
16&News sentiment daily average              & Bloomberg&  Technical  \\
17&ATR                                                            & Bloomberg&  Technical  \\
18&Fear \& Greed                                           &Bloomberg&    Technical \\
19&Hurst                                                     &Bloomberg&    Technical \\
20&Volume                                                   &Bloomberg&    Technical  \\
\end{tabular}
}
\caption{Factors used to regress the standardised innovations obtained with the time-series model. The numerical index is used to identify each factor on the axis of Figure \ref{fig:Pval}.}
\label{tab:factor_list}
\end{table}

The third class is composed of five popular technical analysis indicators. 
The first is the average true range (\textit{ATR}).
It is a measure of the volatility based on the true range indicator, which in turn is the greatest of the following values: the current high minus the current low, the absolute value of the current high minus the previous close, and the absolute value of the current low minus the previous close.
The average true range used in this work is the 14-day  moving average of the true ranges.
The second indicator is the Fear \& Greed index, which is computed as the spread of two weighted moving averages of the true range, so that it oscillates around a zero baseline.
The Fear \& Greed index is used as a measure of the ratio of the buying strength to the selling strength.
The third variable used by technical analysts is the Hurst exponent (HE), which was introduced by Hurst in \cite{Hurst_1959}; he used it to study the Nile's yearly flooding.
It was made popular in economics by Mandelbrot (see, for instance, \cite{Mandelbrot_1968, Lo_1991}).
The HE is a statistic that is extensively used to detect long-range trends in time series. 
The daily news sentiment indicator extrapolates the sentiment of news published during the previous 24 hours on a given stock, providing a number in the interval $\left[ -1 , 1 \right ]$, where a negative (positive) number indicates a pessimistic (optimistic) view of the stock price's movement the next day. The last variable that we added to the factor model is the daily volume.\\

Of course, the factors considered in this study do not cover all of the market information and represent a very limited subset of the three categories; nevertheless, taken together, they form an informative set that is capable of improving the out-of-sample performance, as we will show in the next section.
Since the factors have very different scales, they have been normalised in the interval $\left[0,1\right]$.
The proposed taxonomy of factors in three categories could appear arbitrary, as momentum indicators are extensively used in technical analysis.
However, momentum strategies, as pointed out in the introduction, cover an important class of studies and deserve to be analysed on their own; this will be relevant in the next section, as we will test the effect of the factors on the portfolio performance by first considering all the factors and then by considering only one category of factors at a time.

\begin{figure}[!h]
	\centering
	\subcaptionbox{ adj-$R^{2}$}{\includegraphics[width=.49\textwidth]{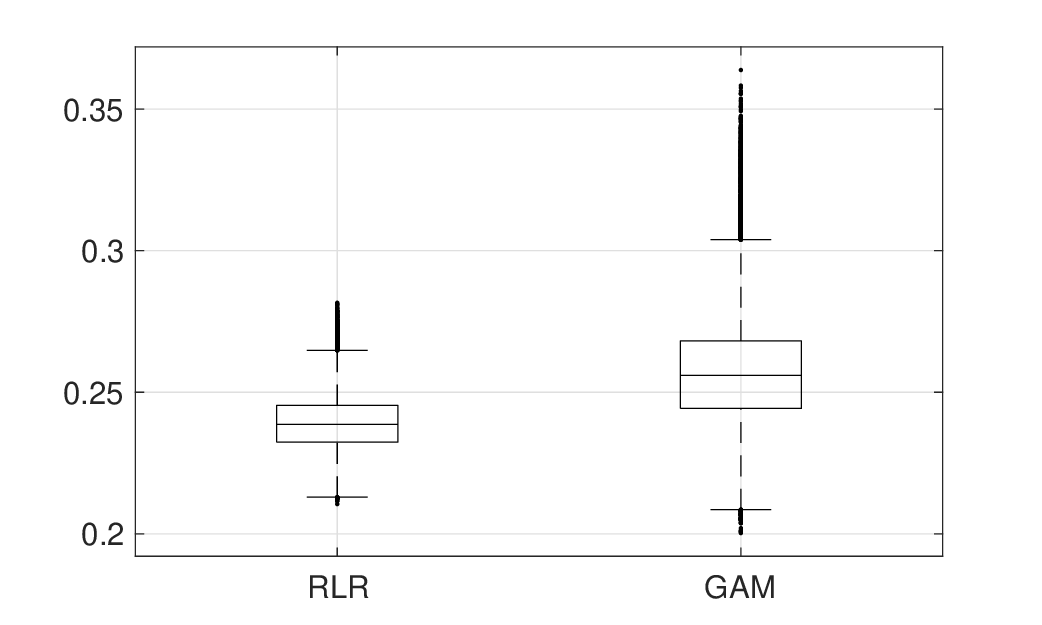} }	
	\subcaptionbox{MAE}{\includegraphics[width=.49\textwidth]{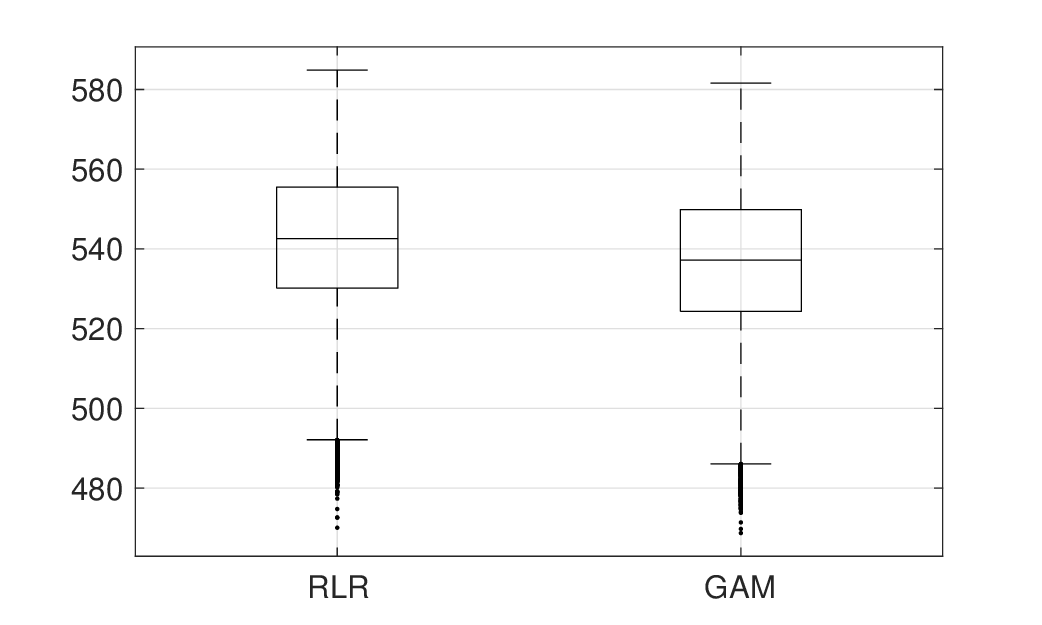}} 	
	
	\vspace{0.2cm}	
	
	\subcaptionbox{BIC}{\includegraphics[width=.49\textwidth]{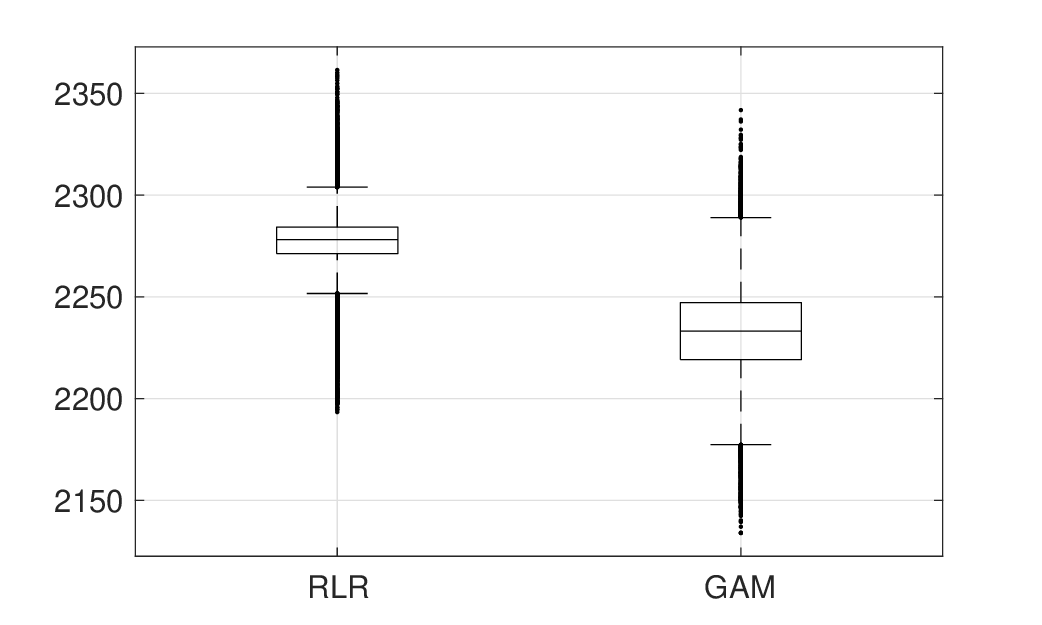} }	
	\caption{ \small Each vignette shows a comparison between the boxplots of a particular statistic for the RLR and GAM, respectively. 
	                         The distributions are obtained by considering all the daily estimations from 2016-05-01 to 2022-04-28 for all the factors and for all the assets.
	                         The statistics are the adjusted $R^{2}$ value, the mean absolute error, and the BIC information criterion, respectively. 
	                         The GAM regression provides better distributions with respect to all three statistics. } 
	\label{fig:BP_RLR_GAM}	
\end{figure}

\begin{figure}[!h]
	\centering
	\includegraphics[width=1\textwidth]{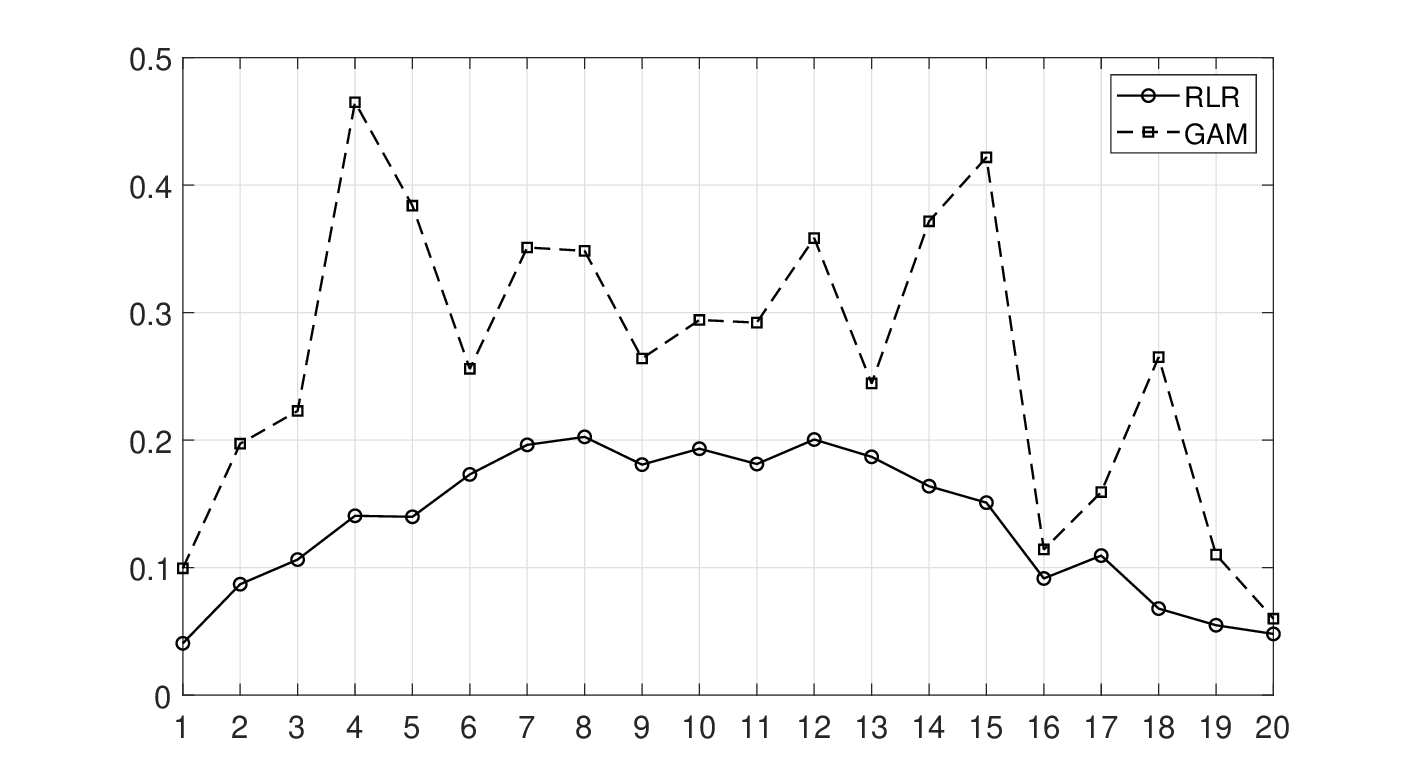}
	\caption{ \small Proportion of the regressions with $p$-values lower than the 0.05 significance level for each of the 20 factors represented on the $x$-axis. 
	The circles connected by the solid line represent the  RLR case, and the squares connected by the dashed line represent the GAM regression.} 
	\label{fig:Pval}	
\end{figure}

\begin{figure}[!h]
	\centering
	\subcaptionbox{DJ30 }{\includegraphics[width=.49\textwidth]{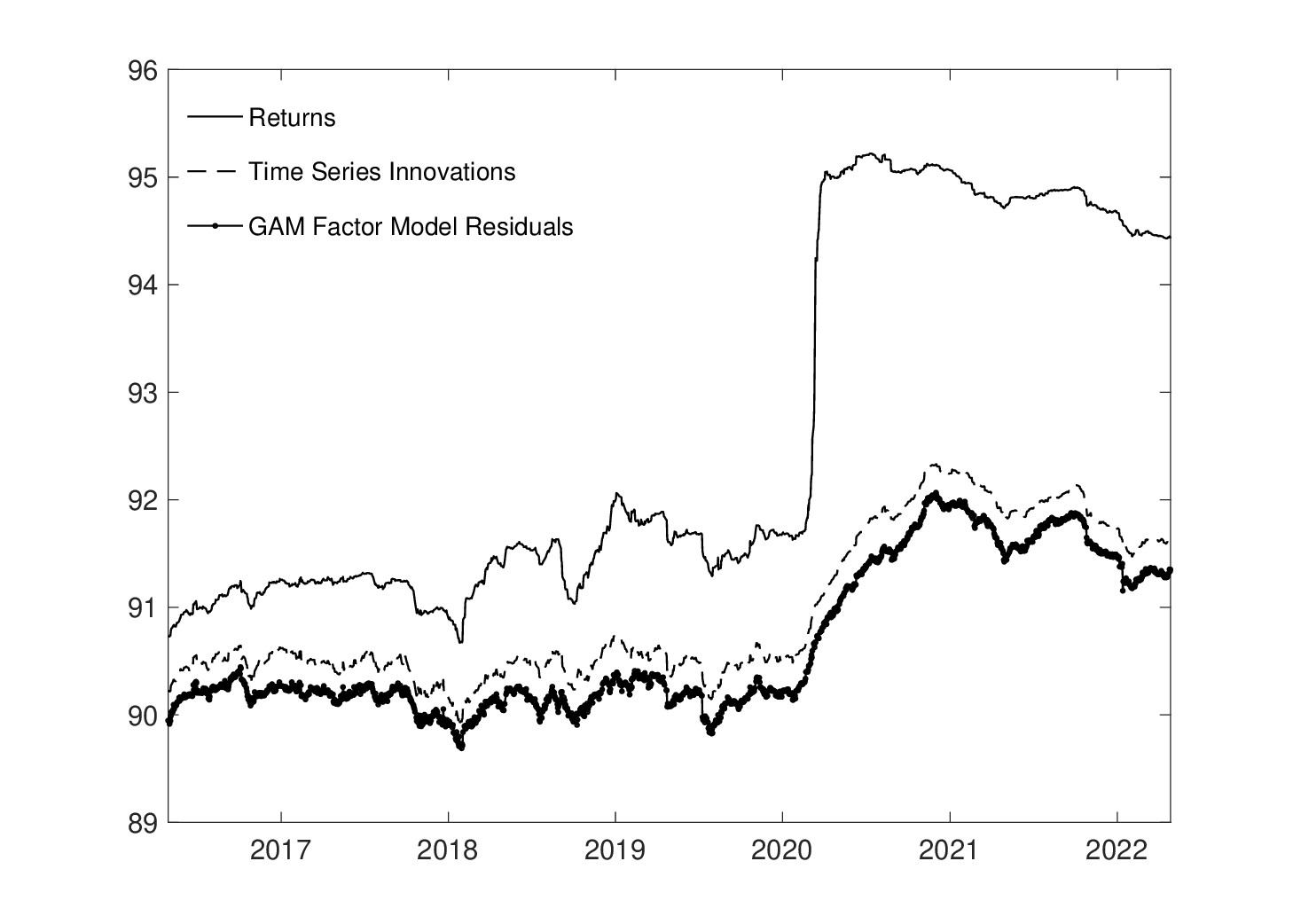} }	
	\subcaptionbox{S\&P500}{\includegraphics[width=.49\textwidth]{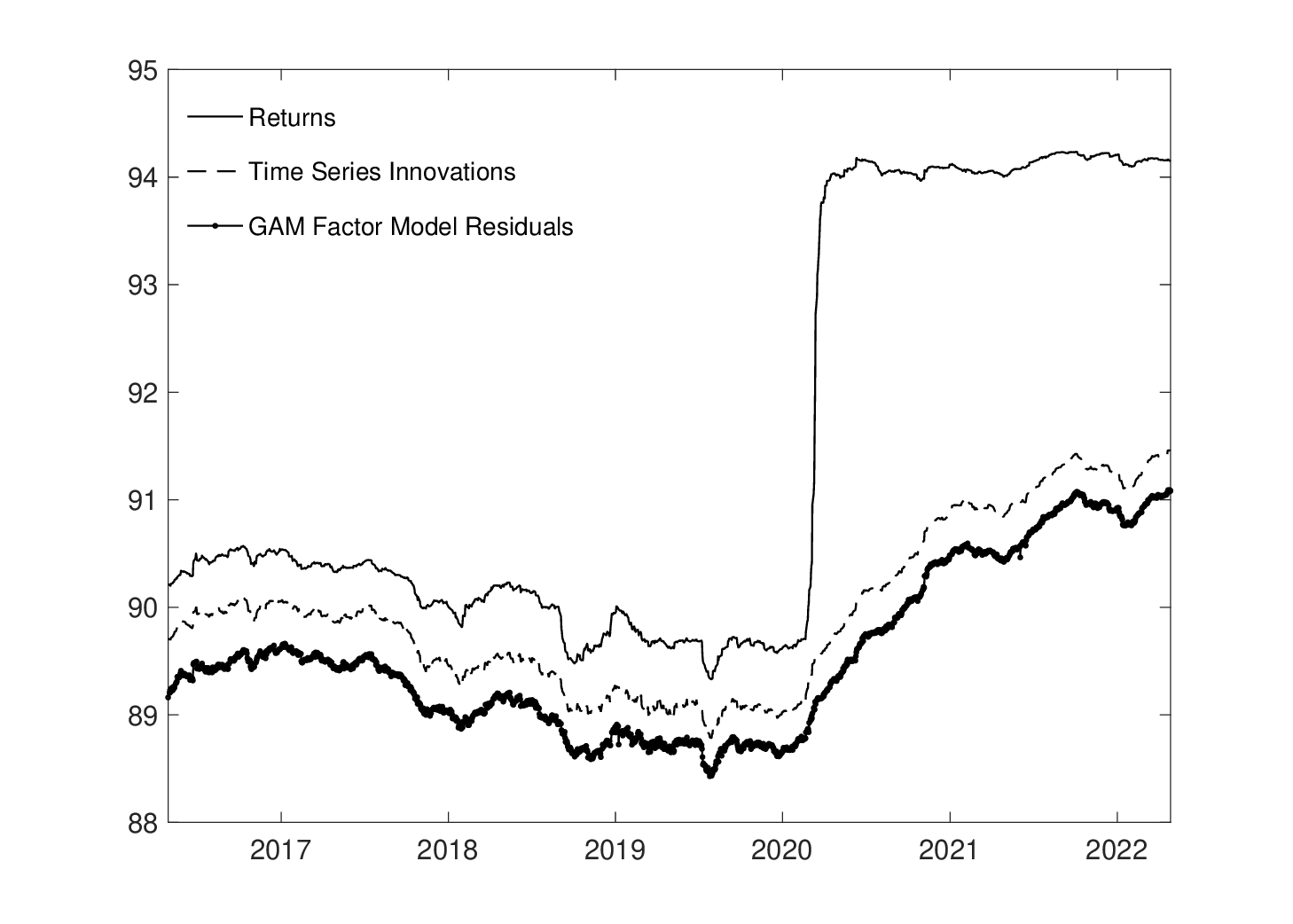}} 	
		
	\caption{ \small Each vignette shows the dynamic of the percentage of the variance explained by the number 
	of principal components that explains at least 90\% of the variance, if running the PCA on row returns at the first date, when the eigenvalues are computed on i) row simple returns (solid line), ii) the innovations of the ARMA-GARCH model (dashed line), iii) the residual of the GAM type regression (solid line with circle marker).} 
	\label{fig:Loadings_90_gam}	
\end{figure}

\section{Out-of-Sample Analysis}\label{sec:EmpiricalAnalysis}  

In this section, we present  results on out-of-sample optimal portfolios obtained by applying three different specifications of the econometric model in order to  to generate the set of $S$ simulations that serve as the input in the optimisation problem. The simplest model does not consider any factors, so it is essentially based solely on the autoregressive ARMA-GARCH filter. The second model and the third model add the RLR model and GAM, respectively, to incorporate information from the factors presented in the previous section. \\

We consider the investment period between 1-May-2016 and 28-April-2022 in the asset universe of the  Dow Jones Industrial Average (DJIA) index, with the exception of the asset \textit{Dow}, as it was recently added to the index and thus its price history does not cover the whole horizon.
The main statistics for the compounded daily returns for each of the 29 assets in the investment universe are collected in Table \ref{tab:hist_ret}.
For each trading day, we estimate the model for a rolling window of $T_{w}=765$ days, generate $S=10000$ one-step-ahead scenarios for asset returns, and run the optimisation problem defined in equations (\ref{eq:MCVeR_objf})--(\ref{eq:MCVaR_const_4}). We consider an equal cost of two basis points for each transaction in order to take into account re-balancing fees that may heavily affect the performance of the portfolio; for the same purpose, we also impose a portfolio turnover constraint of $5\%$ and rule out short-selling positions in order to limit additional complexities, such as capital margins and regulation requirements.\\

In the two cases in which the factor model is present, the two regressions are estimated at each trading date and for each of the 29 assets. 
It is then complex to present  statistics, such as $p$-values and $R^{2}$ coefficients, for each of these regressions. In order to obtain an overall idea of the difference between the RLR and GAM estimates, we have computed for each model the distribution of the $R^{2}$ coefficient, the mean absolute error (MAE), and the BIC information criterion for all the regressions, for each day, and for all the assets.
Each vignette of Figure \ref{fig:BP_RLR_GAM} compares the  boxplots of the empirical distribution of a particular statistic for the two models.
It is clear that not only the location but also the mass of the distribution of $R^{2}$ of the GAM is shifted toward the right with respect to the RLR case, which suggests that the GAM factor model has a greater explanatory power. The distributions of the MAE and the BIC information criterion confirm that the GAM regression outperforms the RLR.
 
Figure \ref{fig:Pval} shows, for each of the twenty factors, the proportion of regressions, for all the assets and all the trading days, for which the $p$-value is lower than 0.05. The circles connected by the solid line represent this proportion for the RLR case, while the squares connected by the dashed line represent the GAM regressions.  The generalised additive model obtained a substantially higher proportion of significative regressions for all the factors, and  the momentum indicator RSI with a 30-day interval achieved the highest score. For both models, the fundamental factors show a higher proportion of significant regressions overall. 

The amount of variance explained by the time series model and the GAM factor model, respectively, can be computed by mean of Principal Components Analysis (PCA). 
At the first date of the out-of-sample analysis we can compute the number of principal components that explain at least 90\% of the variance of the row simple returns on the time windows of three years we used to fit the model. We can now show the dynamics of the explained variance by this number of factors on the entire investment horizon of six years, by computing the PCA on a rolling time windows of three years. We can also compute the  dynamics of the explained variance by this number $k$ of factors on the innovations of the ARMA-GARCH model and the residuals of the GAM type regression, respectively.
The two vignettes in Figure \ref{fig:Loadings_90_gam} show this set of dynamics for the two cases, the DJ30  and the reduced SP\&500.  As expected, the same number of eigenvalues in the factor model explained less variance in the residuals with respect to the case where only the time series filter is applied, suggesting a reduction in the information not captured by the factor model.

\begin{figure}[!h]
	\centering
	\includegraphics[width=1.0\textwidth]{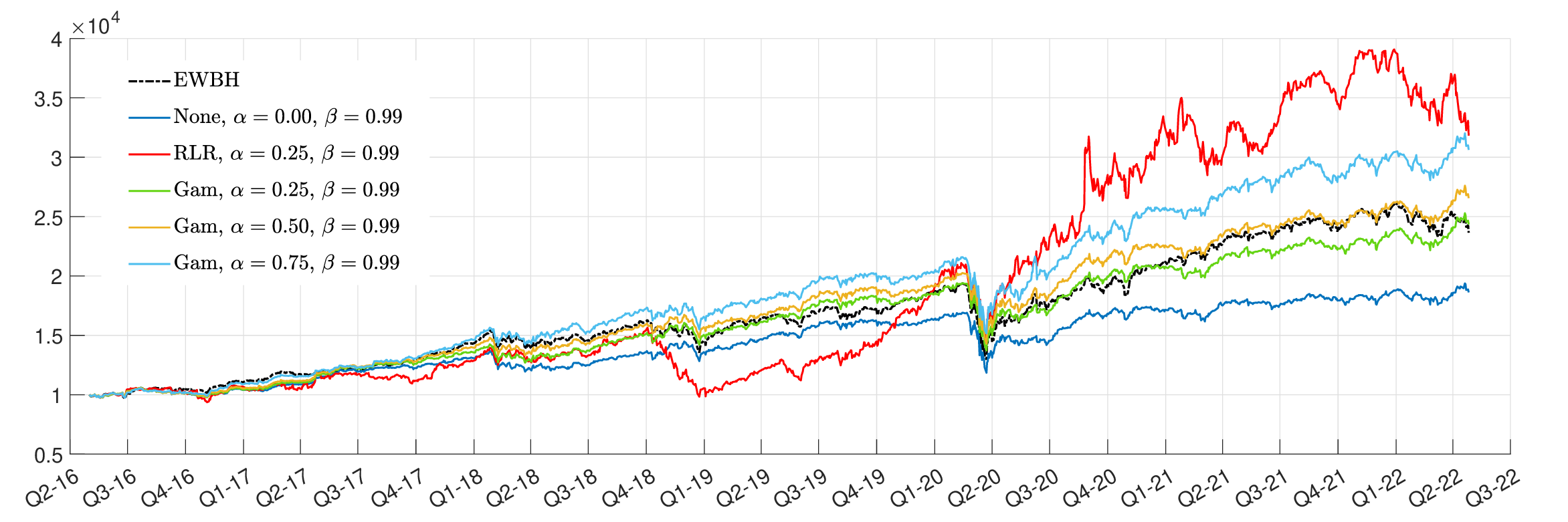} 	
	\caption{ \small Out-of-sample portfolio values from 2016-05-01 to 2022-04-28 for the DJIA case.} 
	\label{fig:PortfolioValues}	
\end{figure}

A comparison of the results of the optimal strategies obtained with the three different models also shows an increase in the out-of-sample performance when the GAM regression is used. Table \ref{tab:OP_Performance} contains summary statistics such as the total and annualised out-of-sample returns, the average turnover (AvgT), and the maximum drawdowns (MD) for each optimal strategy. In order to quantify the extreme realisations, both in terms of extreme losses and extreme gains, we define the lower and upper conditional values-at-risk. Formally, given a random variable $X$ with a distribution function $F_{X}$, we define the $\beta$ lower and upper CVaRs as follows:
\begin{align}	
&\text{CVaR}^{(l)}_{\beta}(X)  := \frac{1}{1-\beta} \int_{ x <  F_{X}^{-1}( 1-\beta ) }x dF_{X},   \label{eq:CvaR_l} \\ 
&\text{CVaR}^{(u)}_{\beta}(X) := \frac{1}{\beta} \int_{ x >  F_{X}^{-1}( \beta) }x dF_{X},    \label{eq:CvaR_u} 
\end{align}
respectively. This definition is the same as the usual definition of the CVaR given by Rockafellar and Uryasev \cite{Rockafellar_2000}, although in this work, the conditional distribution is computed on the returns, not the losses, and the positive part of the tail is also considered.
%
Strategies enhanced with factor regressions obtained higher realised returns compared to the case with only the time-series filter for every tested value of the risk attitude parameter $\alpha$. In the case of the RLR model, the higher total returns come at the price of a higher volatility and tail risk, as the CVaR and MD are considerably higher than they are in the two other cases. The higher volatility of RLR-based strategies can be also verified by looking at Table \ref{tab:OP_MainStat}, which contains the main statistics for the out-of-sample daily returns. In contrast, the strategies based on the GAM factor model obtained higher returns without worsening the risk.\\

A practical and common approach that can be used to measure the trade-off between the expected return and risk, and to compare different strategies, is computing reward-to-risk (RR) ratios.
Table \ref{tab:OP_RR} shows the main RR ratios, such as the information ratio, which is simply the ratio between the sample mean and the variance.
The Sortino ratio is defined as  the average return divided by a lower partial moment of the portfolio return distribution, whereas the Rachev ratio focuses on both positive and negative extreme realisations by computing the ratio between the lower and upper CVaRs, which are defined in equations  (\ref{eq:CvaR_l})--(\ref{eq:CvaR_u}).
We have also considered mean-risk ratios with coherent risk measures: the stable tail adjusted return (STAR), which is the ratio between the expected value and the conditional value-at-risk, and Gini's ratio, as defined by Shalit and Yitzhaki \cite{Shalit_1984}, which is the ratio between expected returns and Gini's mean difference. The last quantity is a risk measure that focuses on the variability of the returns among themselves, instead of considering deviations from a particular central value. 
Among the listed reward-to-risk ratios, the Sortino, STAR, and Gini's ratios  satisfy the four properties of monotonicity, quasi-concavity, scale invariance, and  being based on a distribution presented by Cheridito and Kromer \cite{Cheridito_2013}. \\

The horse race on a selected set of models, depicted in Figure \ref{fig:PortfolioValues}, provides a good visual illustration of the results in terms of  reward-to-risk ratios. We can see that, although the strategy obtained with the RLR model with $\alpha=0.25$ attained the best final return, the portfolio value widely oscillates, with long-lasting periods, for instance, during 2019, of underperformance with respect to the benchmark. 
In order to test if all the factor groups are significant for the out-of-sample portfolio performance,  we consider only one of the three factor groups at a time, and we replicate the econometric model with the GAM regression and portfolio optimisation over the whole period. The results, which are collected in Table \ref{tab:OP_FK}, show that considering only one factor group leads to lower out-of-sample performances with respect to the case in which all the factors are used in the GAM regression.\\  
 
The second empirical test that was used to compare the three models was performed on a  subset of securities belonging to the Standard and Poor's 500 (S\&P500) index.
The large size of the S\&P500 makes it challenging, due to the computational burden, to run the time series and the factor model for each security in the index. Furthermore, data collection for the factors is not possible for all the assets. We proposed a method to reduce the number of assets to avoid some form of biased selection: we wish to create a subset of securities that represent the interval of capitalisation values spanned by the assets in the index. 
We first eliminate the assets for which we do not have enough stock price data in the time interval from May 1, 2013 to April 28, 2022, in order to be consistent with the example performed on the DJIA; this procedure shrinks the number of assets from 500 to 313. Then, instead of considering all the resulting assets, we further reduce the number of securities by  clustering the companies in 100 subsets with respect to their capitalisation,  and then, we select the asset whose capitalisation value is the median of each cluster. 
We discard the companies for which we do not have all the values for the factors, obtaining 96 stocks that represent the different levels of capitalisation in the S\&P500 index.
The tickers and main statistics of the resulting investing universe, over the period from May 1, 2016 to April 28, 2022, are shown in Tables \ref{tab:hist_ret_SP500A}--\ref{tab:hist_ret_SP500B}.
The results for the out-of-sample optimal strategies for the three different models and for a limited selection of the values of the risk parameter $\alpha$ are given in 
Tables \ref{tab:OP_Performance_SP500}--\ref{tab:OP_RR_SP500}. Optimal strategies based on the GAM factor regression outperform all the other strategies for any particular choice of the risk attitude parameter $\alpha$. Contrary to what happened in the DJIA case, the RLR-based strategies achieved the worst results, both in terms of total returns and RR ratios, confirming the higher volatility of the out-of-sample results for these strategies compared to the other two approaches that we have tested.

\section{Discussion} \label{sec:Disc}

In this work, we have empirically tested the advantages of combining time-series filters and regression models in practical portfolio optimisation. 
Such advantages have been measured using statistical properties, and five popular reward-to-risk ratios of out-of-sample returns obtained by implementing mean-CVaR optimal portfolio strategies, for a period of six years, on two different asset universes from the DJIA and S\&P500 indexes, respectively, were calculated.

The results show that GAM factor models, in general, increase the out-of-sample performance of the optimal portfolio strategy compared to the case in which only the autoregression time-series model is used to predict return scenarios. 
Between the two alternative factor models, namely the RLR and GAM, that we have tested, the latter obtained the best risk-to-reward ratios for all the possible specifications of the risk parameter $\alpha$ in the portfolio optimisation procedure. The empirical results of the analysis suggest that the extra information from GAM factor models improved the RRs of mean-CVaR optimal portfolios, while the same information, if processed with a standard robust linear regression method, increases the variability of each strategy.  
The positive results obtained with the GAM factor model suggest that this approach could be applied to other asset universes to provide similar improvements compared to the case in which only time-series models are implemented. 
Two extensions of this research naturally arise. The first would involve testing the GAM models against other models such as neural networks, support vector machines, and generalised linear models. The second would involve including other sets of factors that could provide more informative signals.

\newpage
\begin{appendices}

\renewcommand{\thetable}{A.\arabic{table}}
\setcounter{table}{0}
\section{Tables} \label{sec:Tbl}
\begin{table}[h!] 
	\centering
	\scalebox{0.85}{
	\begin{tabular}{ lrrrrrrr } \hline \\ 
		Ticker  & Mean & Median & Std & Skewness & Kurtosis & CVaR$^{(l)}_{0.99}$ & CVaR$^{(u)}_{0.99}$  \\ \hline \\ 
	MMM & 0.0001 & 0.0006 & 0.015 & $-0.572$ & 14.470 & $-0.067$ & 0.057  \\  
	AXP & 0.0009 & 0.0010 & 0.021 & 1.445 & 27.672 & $-0.086$ & 0.108  \\  
	AMGN & 0.0005 & 0.0006 & 0.016 & 0.262 & 10.324 & $-0.061$ & 0.067  \\  
	AAPL & 0.0015 & 0.0010 & 0.018 & $-0.026$ & 9.633 & $-0.071$ & 0.076  \\  
	BA & 0.0006 & 0.0003 & 0.028 & 0.332 & 20.330 & $-0.121$ & 0.142  \\  
	CAT & 0.0010 & 0.0007 & 0.019 & $-0.338$ & 7.987 & $-0.077$ & 0.067  \\  
	CVX & 0.0007 & 0.0007 & 0.020 & $-0.302$ & 32.848 & $-0.086$ & 0.087  \\  
	CSCO & 0.0006 & 0.0009 & 0.016 &$ -0.238$ & 13.505 & $-0.071$ & 0.066  \\  
	KO & 0.0005 & 0.0007 & 0.012 & -0.773 & 13.846 & $-0.058$ & 0.050  \\  
	GS & 0.0007 & 0.0004 & 0.019 & 0.163 & 14.283 & $-0.081$ & 0.081  \\  
	HD & 0.0008 & 0.0010 & 0.016 & $-1.288$ & 27.614 & $-0.073$ & 0.065  \\  
	HON & 0.0006 & 0.0010 & 0.015 & 0.037 & 17.871 & $-0.067$ & 0.065  \\  
	IBM & 0.0003 & 0.0005 & 0.016 & $-0.457$ & 14.005 & $-0.074$ & 0.066  \\  
	INTC & 0.0006 & 0.0008 & 0.021 &$ -0.329$ & 17.830 & $-0.095$ & 0.087  \\  
	JNJ & 0.0005 & 0.0005 & 0.012 & $-0.354$ & 14.011 & $-0.055$ & 0.052  \\  
	JPM & 0.0007 & 0.0003 & 0.018 & 0.332 & 18.981 & $-0.075$ & 0.082  \\  
	MCD & 0.0006 & 0.0008 & 0.014 & 0.480 & 39.523 & $-0.062$ & 0.064  \\  
	MRK & 0.0005 & 0.0002 & 0.014 & 0.117 & 11.843 & $-0.057$ & 0.062  \\  
	MSFT & 0.0014 & 0.0011 & 0.017 & 0.002 & 12.851 & $-0.063$ & 0.072  \\  
	NKE & 0.0007 & 0.0004 & 0.018 & 0.738 & 15.294 & $-0.068$ & 0.083  \\  
	PG & 0.0006 & 0.0008 & 0.012 & 0.489 & 18.102 & $-0.052$ & 0.058  \\  
	CRM & 0.0008 & 0.0009 & 0.021 & 0.753 & 22.238 & $-0.080$ & 0.085  \\  
	TRV & 0.0005 & 0.0011 & 0.016 & $-1.785$ & 31.282 & $-0.077$ & 0.061  \\  
	UNH & 0.0011 & 0.0010 & 0.017 &$ -0.047$ & 18.336 & $-0.069$ & 0.082  \\  
	VZ & 0.0002 & 0.0002 & 0.012 & 0.243 & 9.032 & $-0.043$ & 0.049  \\  
	V & 0.0008 & 0.0015 & 0.017 & 0.208 & 14.246 & $-0.064$ & 0.072  \\  
	WBA & $-0.0001$ & 0.0002 & 0.019 & $-0.239$ & 9.904 & $-0.080$ & 0.072  \\  
	WMT & 0.0007 & 0.0006 & 0.013 & 1.055 & 19.567 & $-0.050$ & 0.071  \\  
	DIS & 0.0003 & 0.0000 & 0.018 & 0.779 & 16.759 & $-0.070$ & 0.086  \\  
		\\ \hline \\  
	\end{tabular} 
}
\caption{ \small Main statistics for logarithmic returns of 29 assets in the DJIA index computed on the sample from 2016-05-01 to 2022-04-28.}
\label{tab:hist_ret}
\end{table} 

\begin{table}[] 
	\centering
	\scalebox{0.70}{
 \begin{tabular}{ lrrrrrrr } \hline \\ 
Ticker  & Mean & Median & Std & Skewness & Kurtosis & CVaR$^{(l)}_{0.99}$ & CVaR$^{(u)}_{0.99}$  \\ \hline \\ 
RL       & 0.0005 & 0.0009 & 0.025 & 0.289 & 10.352 & $-0.097$ & 0.107  \\  
AIZ      & 0.0008 & 0.0008 & 0.017 & $-0.929$ & 24.308 & $-0.078$ & 0.073  \\  
MOS   & 0.0011 & 0.0009 & 0.030 & $-0.245$ & 12.054 & $-0.116$ & 0.119  \\  
PNW   & 0.0003 & 0.0009 & 0.015 & $-0.761$ & 23.608 & $-0.067$ & 0.065  \\  
SNA    & 0.0005 & 0.0009 & 0.018 & 0.099 & 9.900 & $-0.075$ & 0.079  \\  
TAP     & $-0.0001$ & 0.0000 & 0.019 & $-0.433$ & 10.193 & $-0.080$ & 0.071  \\  
HST    & 0.0006 & 0.0006 & 0.023 & 1.680 & 27.336 & $-0.085$ & 0.109  \\  
UDR   & 0.0006 & 0.0013 & 0.017 & $-0.223$ & 19.592 & $-0.071$ & 0.075  \\  
IFF     & 0.0003 & 0.0004 & 0.018 & $-0.229$ & 16.498 & $-0.080$ & 0.074  \\  
HAS   & 0.0004 & 0.0006 & 0.021 & 0.326 & 24.301 & $-0.096$ & 0.104  \\  
AVY   & 0.0008 & 0.0006 & 0.018 & 1.098 & 25.748 & $-0.068$ & 0.078  \\  
EMN  & 0.0005 & 0.0008 & 0.019 & $-0.395$ & 10.675 & $-0.081$ & 0.072  \\  
J        & 0.0010 & 0.0005 & 0.018 & 0.038 & 9.811 & $-0.070$ & 0.077  \\  
CPB   & 0.0001 & 0.0006 & 0.016 & $-0.286$ & 11.842 & $-0.073$ & 0.066  \\  
FMC  & 0.0011 & 0.0011 & 0.021 & 0.061 & 16.391 & $-0.082$ & 0.098  \\  
CE     & 0.0007 & 0.0009 & 0.020 & $-0.790$ & 12.966 & $-0.086$ & 0.077  \\  
RF     & 0.0010 & 0.0009 & 0.025 & $-0.296$ & 12.742 & $-0.101$ & 0.101  \\  
AES  & 0.0008 & 0.0010 & 0.021 & $-0.651$ & 14.713 & $-0.094$ & 0.083  \\  
STX  & 0.0012 & 0.0020 & 0.025 & $-0.009$ & 13.237 & $-0.097$ & 0.100  \\  
NUE & 0.0011 & 0.0006 & 0.022 & $-0.050$ & 7.747 & $-0.085$ & 0.084  \\  
EXPD & 0.0006 & 0.0010 & 0.015 & 0.060 & 11.025 & $-0.056$ & 0.055  \\  
PEAK & 0.0004 & 0.0013 & 0.019 & $-0.907$ & 25.136 & $-0.086$ & 0.076  \\  
HAL & 0.0005 & $-0.0004$ & 0.031 & $-0.844$ & 26.145 & $-0.135$ & 0.132  \\  
AKAM & 0.0007 & 0.0016 & 0.019 & $-0.341$ & 18.902 & $-0.092$ & 0.080  \\  
MLM & 0.0007 & 0.0005 & 0.021 & 0.137 & 10.138 & $-0.082$ & 0.089  \\  
COO & 0.0007 & 0.0014 & 0.016 & $-0.726$ & 12.906 & $-0.073$ & 0.063  \\  
EXPE & 0.0007 & 0.0007 & 0.026 & $-0.508$ & 25.623 & $-0.128$ & 0.116  \\  
NTRS & 0.0006 & 0.0007 & 0.019 & 0.157 & 19.448 & $-0.075$ & 0.076  \\  
VMC & 0.0006 & 0.0000 & 0.021 & 0.399 & 16.179 & $-0.079$ & 0.089  \\  
ETR & 0.0006 & 0.0009 & 0.016 & $-0.386$ & 24.495 & $-0.070$ & 0.065  \\  
K & 0.0002 & 0.0006 & 0.014 & $-0.067$ & 13.681 & $-0.058$ & 0.060  \\  
GWW & 0.0008 & 0.0009 & 0.019 & 0.484 & 17.554 & $-0.079$ & 0.086  \\  
CCL & 0.0001 & 0.0004 & 0.037 & $-0.060$ & 25.385 & $-0.173$ & 0.171  \\  
VLO & 0.0009 & 0.0009 & 0.027 & 0.961 & 22.144 & $-0.104$ & 0.124  \\  
DTE & 0.0006 & 0.0011 & 0.015 & $-0.726$ & 25.010 & $-0.073$ & 0.071  \\  
KR & 0.0006 & 0.0009 & 0.020 &$ -0.625$ & 14.120 & $-0.087$ & 0.076  \\  
WY & 0.0006 & 0.0011 & 0.022 & $-0.071$ & 31.138 & $-0.103$ & 0.097  \\  
CLX & 0.0003 & 0.0011 & 0.015 & $-0.859$ & 18.208 & $-0.071$ & 0.055  \\  
BBY & 0.0011 & 0.0019 & 0.024 & 0.726 & 15.700 & $-0.094$ & 0.117  \\  
LUV & 0.0003 & 0.0002 & 0.023 & 0.017 & 9.887 & $-0.096$ & 0.093  \\  
FAST & 0.0008 & 0.0011 & 0.017 & 0.680 & 14.406 & $-0.064$ & 0.079  \\  
MSI & 0.0009 & 0.0013 & 0.017 & 0.116 & 14.075 & $-0.070$ & 0.069  \\  
ROK & 0.0008 & 0.0009 & 0.019 & 0.201 & 18.005 & $-0.072$ & 0.086  \\  
SLB & 0.0000 & $-0.0006$ & 0.027 & $-0.279$ & 16.414 & $-0.110$ & 0.111  \\  
PRU & 0.0007 & 0.0010 & 0.022 & $-0.361$ & 21.261 & $-0.100$ & 0.093  \\  
HPQ & 0.0011 & 0.0016 & 0.022 & $-0.493$ & 15.072 & $-0.099$ & 0.090  \\  
YUM & 0.0007 & 0.0008 & 0.015 & 1.962 & 45.108 & $-0.063$ & 0.067  \\  
PAYX & 0.0009 & 0.0011 & 0.016 & $-0.440$ & 32.488 & $-0.071$ & 0.072  \\  
TROW & 0.0007 & 0.0011 & 0.018 & 0.115 & 15.017 & $-0.073$ & 0.075  \\  
 
 \\ \hline \\  
\end{tabular} 
}
\caption{ \small Main statistics for logarithmic returns of the assets belonging to the SP\&500 index computed on the sample from 2016-05-01 to 2022-04-28.}
\label{tab:hist_ret_SP500A}
\end{table} 


\begin{table}[] 
	\centering
	\scalebox{0.70}{
 \begin{tabular}{ lrrrrrrr } \hline \\ 
Ticker  & Mean & Median & Std & Skewness & Kurtosis & CVaR$^{(l)}_{0.99}$ & CVaR$^{(u)}_{0.99}$  \\ \hline \\ 

EBAY & 0.0007 & 0.0009 & 0.019 & $-0.089$ & 9.311 & $-0.074$ & 0.071  \\  
XEL & 0.0006 & 0.0009 & 0.014 & $-0.491$ & 19.321 & $-0.057$ & 0.059  \\  
GIS & 0.0003 & 0.0007 & 0.014 & $-0.452$ & 12.237 & $-0.060$ & 0.049  \\  
SRE & 0.0006 & 0.0008 & 0.016 & $-0.354$ & 28.563 & $-0.071$ & 0.074  \\  
BIIB & 0.0002 & 0.0000 & 0.028 & 3.853 & 81.963 & $-0.106$ & 0.155  \\  
CDNS & 0.0015 & 0.0019 & 0.020 & 0.295 & 10.199 & $-0.073$ & 0.087  \\  
LHX & 0.0009 & 0.0011 & 0.017 & 0.088 & 14.455 & $-0.064$ & 0.078  \\  
BAX & 0.0005 & 0.0011 & 0.015 & $-0.474$ & 13.933 & $-0.065$ & 0.057  \\  
EA & 0.0006 & 0.0006 & 0.019 & 0.458 & 12.443 & $-0.070$ & 0.083  \\  
COP & 0.0009 & 0.0000 & 0.026 & 0.146 & 19.303 & $-0.100$ & 0.112  \\  
CTSH & 0.0004 & 0.0008 & 0.018 & $-0.432$ & 18.391 & $-0.084$ & 0.077  \\  
LVS & 0.0003 & 0.0001 & 0.025 & $-0.007$ & 9.091 & $-0.098$ & 0.097  \\  
EMR & 0.0006 & 0.0008 & 0.019 & $-0.554$ & 20.118 & $-0.079$ & 0.077  \\  
WM & 0.0009 & 0.0011 & 0.013 & $-0.395$ & 17.372 & $-0.055$ & 0.056  \\  
HUM & 0.0008 & 0.0006 & 0.020 & $-0.174$ & 19.217 & $-0.087$ & 0.088  \\  
PGR & 0.0010 & 0.0011 & 0.015 & 0.089 & 10.898 & $-0.061$ & 0.063  \\  
APD & 0.0006 & 0.0009 & 0.016 & $-0.192$ & 16.173 & $-0.069$ & 0.065  \\  
ECL & 0.0004 & 0.0009 & 0.016 & 0.810 & 34.908 & $-0.075$ & 0.071  \\  
TFC & 0.0006 & 0.0008 & 0.022 & $-0.149$ & 16.795 & $-0.092$ & 0.097  \\  
SHW & 0.0009 & 0.0009 & 0.017 & $-0.433$ & 22.365 & -$0.070$ & 0.072  \\  
LRCX & 0.0016 & 0.0022 & 0.027 & 0.125 & 10.150 & $-0.097$ & 0.106  \\  
CSX & 0.0011 & 0.0007 & 0.019 & 1.027 & 26.778 & $-0.078$ & 0.086  \\  
CL & 0.0003 & 0.0004 & 0.013 & 0.376 & 17.120 & $-0.052$ & 0.057  \\  
ADP & 0.0008 & 0.0014 & 0.016 & -0.483 & 17.440 & $-0.072$ & 0.070  \\  
AMAT & 0.0015 & 0.0016 & 0.026 & $-0.257$ & 8.294 & $-0.096$ & 0.096  \\  
TJX & 0.0005 & 0.0006 & 0.018 & $-0.469$ & 19.722 & $-0.073$ & 0.077  \\  
DE & 0.0013 & 0.0011 & 0.020 & 0.164 & 10.429 & $-0.073$ & 0.081  \\  
CVS & 0.0003 & 0.0003 & 0.017 & $-0.418$ & 10.577 & $-0.073$ & 0.065  \\  
GE & -0.0004 & -0.0010 & 0.025 & 0.159 & 8.544 & $-0.095$ & 0.098  \\  
CAT & 0.0010 & 0.0007 & 0.019 & $-0.338$ & 7.987 & $-0.077$ & 0.067  \\  
AMT & 0.0008 & 0.0012 & 0.016 & 0.006 & 15.856 & $-0.063$ & 0.070  \\  
RTX & 0.0006 & 0.0007 & 0.019 & 0.176 & 18.277 & $-0.082$ & 0.085  \\  
LOW & 0.0009 & 0.0012 & 0.019 & $-1.280$ & 27.804 & $-0.088$ & 0.083  \\  
WFC & 0.0003 & 0.0002 & 0.021 & $-0.015$ & 12.145 & $-0.088$ & 0.092  \\  
AMGN & 0.0005 & 0.0006 & 0.016 & 0.262 & 10.324 & $-0.061$ & 0.067  \\  
UNP & 0.0009 & 0.0009 & 0.017 & $-0.288$ & 13.352 & $-0.069$ & 0.067  \\  
TXN & 0.0010 & 0.0016 & 0.018 & $-0.069$ & 9.005 & $-0.069$ & 0.073  \\  
DHR & 0.0009 & 0.0008 & 0.015 & 0.040 & 8.851 & $-0.054$ & 0.058  \\  
CVX & 0.0007 & 0.0007 & 0.020 & $-0.302$ & 32.848 & $-0.086$& 0.087  \\  
ACN & 0.0008 & 0.0016 & 0.016 & 0.221 & 11.795 & $-0.061$ & 0.069  \\  
CSCO & 0.0006 & 0.0009 & 0.016 & $-0.238$ & 13.505 & $-0.071$ & 0.066  \\  
INTC & 0.0006 & 0.0008 & 0.021 & $-0.329$ & 17.829 & $-0.095$ & 0.087  \\  
MRK & 0.0005 & 0.0002 & 0.014 & 0.117 & 11.843 & $-0.057$ & 0.062  \\  
ADBE & 0.0012 & 0.0018 & 0.021 & 0.072 & 12.073 & $-0.081$ & 0.088  \\  
HD & 0.0008 & 0.0010 & 0.016 & $-1.288$ & 27.615 & $-0.073$ & 0.065  \\  
UNH & 0.0011 & 0.0010 & 0.017 & $-0.047$ & 18.337 & $-0.069$ & 0.082  \\  
WMT & 0.0007 & 0.0006 & 0.013 & $1.055$ & 19.567 & $-0.050$ & 0.071  \\  
AAPL & 0.0015 & 0.0010 & 0.018 & $-0.026$ & 9.632 & $-0.071$ & 0.076  \\  
 \\ \hline \\  
\end{tabular} 
}
\caption{ \small Main statistics for logarithmic returns of the assets belonging to the SP\&500 index computed on the sample from 2016-05-01 to 2022-04-28.}
\label{tab:hist_ret_SP500B}
\end{table} 


\begin{table}[] 
	\centering
	\scalebox{0.95}{
	\begin{tabular}{ lcccccc } \hline \\ 
		
	 Model Type & Tot. Ret & Ann. Ret & AvgT  & CVaR$^{(l)}_{0.95}$ & CVaR$^{(u)}_{0.95} $&  MD  \\ \hline \\ 
	 
	EWBH                      & 136.49 & 15.67 & 0.00 & -2.98 & 2.51 & 33.27  \\ \hline \\  
	
         none, $\alpha=$ 0.00& 86.23 & 11.09 & 2.71 & $-2.31$ & 2.04 & 29.94  \\ 
	none, $\alpha=$ 0.25 & 85.52 & 11.02 & 2.71 & $-2.31$ & 2.05 & 29.79  \\  
	none, $\alpha=$ 0.50 & 84.18 & 10.88 & 2.71 & $-2.32$ & 2.05 & 30.07  \\   
	none, $\alpha=$ 0.75 & 83.05 & 10.76 & 2.71 & $-2.33$ & 2.06 & 29.76  \\   
	none, $\alpha=$ 0.85 & 76.29 & 10.06 & 2.71 & $-2.36$ & 2.07 & 29.71  \\   
	none, $\alpha=$ 0.90 & 73.23 & 9.74 & 2.72 & $-2.39$ & 2.10 & 29.71  \\  
	none, $\alpha=$ 0.95 & 92.98 & 11.76 & 2.72 & $-2.56$ & 2.29 & 28.95  \\   
	none, $\alpha=$ 0.98 & 129.13 & 15.05 & 2.68 & $-3.28$ & 2.93 & 30.32  \\   \hline \\
		
         RLR, $\alpha=$ 0.00 & 224.45 & 22.02 & 1.90 & $-4.06$ & 3.94 & 36.56  \\
	 RLR, $\alpha=$ 0.25 & 218.26 & 21.62 & 1.85 & $-4.12$ & 3.97 & 37.09  \\  
	 RLR, $\alpha=$ 0.50 & 213.98 & 21.35 & 1.81 & $-4.18$ & 4.00 & 38.08  \\  
	 RLR, $\alpha=$ 0.75 & 209.22 & 21.03 & 1.77 & $-4.22$ & 4.02 & 38.79  \\   
	 RLR, $\alpha=$ 0.85 & 210.14 & 21.09 & 1.77 & $-4.22$ & 4.02 & 38.79  \\   
	 RLR, $\alpha=$ 0.90 & 207.88 & 20.94 & 1.75 & $-4.24$ & 4.04 & 39.26  \\   
	 RLR, $\alpha=$ 0.95 & 206.42 & 20.85 & 1.74 & $-4.25$ & 4.05 & 39.47  \\   
	 RLR, $\alpha=$ 0.98 & 208.39 & 20.98 & 1.74 & $-4.25$ & 4.05 & 39.43  \\ \hline \\ 
	
	GAM, $\alpha=$ 0.00 & 134.00 & 15.46 & 2.72 & $-2.32$ & 2.07 & 29.58  \\ 
	GAM, $\alpha=$ 0.25 & 143.48 & 16.24 & 2.73 & $-2.33$ & 2.09 & 29.23  \\  
	GAM, $\alpha=$ 0.50 & 165.57 & 17.96 & 2.74 & $-2.39$ & 2.17 & 29.34  \\ 
	GAM, $\alpha=$ 0.75 & 206.24 & 20.83 & 2.78 & $-2.65$ & 2.48 & 29.78  \\   
	GAM, $\alpha=$ 0.85 & 233.13 & 22.57 & 2.80 & $-3.04$ & 2.93 & 30.13  \\   
	GAM, $\alpha=$ 0.90 & 245.34 & 23.32 & 2.81 & $-3.30$ & 3.20 & 30.08  \\  
	GAM, $\alpha=$ 0.95 & 248.07 & 23.48 & 2.79 & $-3.69$ & 3.66 & 29.78  \\  
	GAM, $\alpha=$ 0.98 & 231.76 & 22.48 & 2.77 & $-3.82$ & 3.76 & 30.23  \\ \hline \\ 

	\end{tabular} 
}
	\caption{ \small Summary measures of the out-of-sample daily optimal portfolio returns from 1-May-2016 to 28-April-2022.}
	\label{tab:OP_Performance}
\end{table} 

\begin{table}[] 
\centering
\scalebox{0.9}{
	\begin{tabular}{ lcccccc } \hline \\ 
		
		Model Type                & Mean & Median & Std & Skew & ExKurt & SDev  \\ \hline \\ 
		
		EWBH & 0.00057 & 0.00084  & 0.012 & $-0.990$ & 23.850 & 0.009  \\ \hline \\

		none, $\alpha=$ 0.00 & 0.00041 & 0.00079 & 0.009 & $-1.079$ & 22.642 & 0.007  \\   
		none, $\alpha=$ 0.25 & 0.00041 & 0.00078 & 0.010 & $-1.053$ & 22.676 & 0.007  \\   
		none, $\alpha=$ 0.50 & 0.00041 & 0.00074 & 0.010 & $-1.084$ & 22.611 & 0.007  \\  
		none, $\alpha=$ 0.75 & 0.00040 & 0.00075 & 0.010 & $-1.109$ & 23.568 & 0.007  \\   
		none, $\alpha=$ 0.85 & 0.00038 & 0.00068 & 0.010 & $-1.100$ & 22.809 & 0.007  \\   
		none, $\alpha=$ 0.90 & 0.00036 & 0.00062 & 0.010 & $-1.092$ & 22.061 & 0.007  \\ 
		none, $\alpha=$ 0.95 & 0.00044 & 0.00088 & 0.011 & $-1.089$ & 20.576 & 0.008  \\ 
		none, $\alpha=$ 0.98& 0.00055 & 0.00100 & 0.014 & $-0.694$ & 11.854 & 0.010  \\  \hline \\   
		
		 RLR, $\alpha=$ 0.00  & 0.00078 & 0.00097 & 0.017 & 0.202 & 12.478 & 0.012  \\ 
		 RLR, $\alpha=$ 0.25  & 0.00077 & 0.00084 & 0.017 & 0.181 & 12.045 & 0.012  \\   
		 RLR, $\alpha=$ 0.50  & 0.00076 & 0.00073 & 0.017 & 0.155 & 11.820 & 0.012  \\  
		 RLR, $\alpha=$ 0.75  & 0.00075 & 0.00067 & 0.018 & 0.132 & 11.570 & 0.012  \\   
		 RLR, $\alpha=$ 0.85  & 0.00075 & 0.00069 & 0.018 & 0.134 & 11.613 & 0.012  \\   
		 RLR, $\alpha=$ 0.90  & 0.00075 & 0.00069 & 0.018 & 0.116 & 11.459 & 0.012  \\  
		 RLR, $\alpha=$ 0.95  & 0.00074 & 0.00070 & 0.018 & 0.109 & 11.429 & 0.012  \\  
		 RLR, $\alpha=$ 0.98  & 0.00075 & 0.00070 & 0.018 & 0.110 & 11.428 & 0.012  \\ \hline \\ 
		
	        GAM, $\alpha=$ 0.00 & 0.00056 & 0.00077 & 0.010 & $-0.939$ & 20.029 & 0.007  \\
		GAM, $\alpha=$ 0.25  & 0.00059 & 0.00068 & 0.010 & $-0.906$ & 19.407 & 0.007  \\ 
		GAM, $\alpha=$ 0.50  & 0.00065 & 0.00085 & 0.010 & $-0.889$ & 19.355 & 0.007  \\   
		GAM, $\alpha=$ 0.75  & 0.00074 & 0.00090 & 0.011 & $-0.665$ & 16.714 & 0.008  \\   
		GAM, $\alpha=$ 0.85  & 0.00080 & 0.00100 & 0.013 & $-0.567$ & 17.060 & 0.009  \\   
		GAM, $\alpha=$ 0.90  & 0.00082 & 0.00105 & 0.014 & $-0.550$ & 17.391 & 0.010  \\   
		GAM, $\alpha=$ 0.95  & 0.00083 & 0.00105 & 0.015 & $-0.388$ & 16.302 & 0.011  \\   
		GAM, $\alpha=$ 0.98  & 0.00080 & 0.00097 & 0.016 & $-0.364$ & 15.394 & 0.011  \\ \hline \\  
		
		\end{tabular}
		} 
	\caption{ \small Main statistics of the out-of-sample daily optimal portfolio returns from 2016-05-01 to 2022-04-28. Strategies based on the RLR factor model exhibit higher values for the two location statistics at the cost of a substantially higher standard deviation (Std) and standard semi-deviation (SDev).  }
	\label{tab:OP_MainStat}
\end{table} 

\begin{table}[] 
\centering
	\begin{tabular}{ lccccc } \hline \\ 
		
		Model Type & IR & Sortino & STAR & Rachev & Gini  \\ \hline \\ 
		
		EWBH                     & 4.81 & 6.58 & 1.92 & 84.11 & 11.74  \\ \hline \\  
		
		none, $\alpha=$ 0.00 & 4.35 & 5.94 & 1.78 & 88.35 & 10.51  \\  
		none, $\alpha=$ 0.25 & 4.32 & 5.90 & 1.77 & 88.61 & 10.42  \\  
		none, $\alpha=$ 0.50 & 4.26 & 5.82 & 1.75 & 88.39 & 10.27  \\ 
		none, $\alpha=$ 0.75 & 4.19 & 5.72 & 1.72 & 88.20 & 10.10  \\  
		none, $\alpha=$ 0.85 & 3.90 & 5.31 & 1.60 & 87.72 & 9.27  \\   
		none, $\alpha=$ 0.90 & 3.72 & 5.07 & 1.53 & 87.78 & 8.72  \\  
		none, $\alpha=$ 0.95 & 4.11 & 5.62 & 1.70 & 89.53 & 9.43  \\   
		none, $\alpha=$ 0.98 & 4.07 & 5.64 & 1.68 & 89.55 & 8.82  \\  \hline \\ 
		
		 RLR, $\alpha=$ 0.00 & 4.57 & 6.61 & 1.92 & 96.93 & 10.20  \\ 
		 RLR, $\alpha=$ 0.25 & 4.45 & 6.42 & 1.86 & 96.44 & 9.88  \\   
		 RLR, $\alpha=$ 0.50 & 4.36 & 6.28 & 1.82 & 95.83 & 9.65  \\   
		 RLR, $\alpha=$ 0.75 & 4.27 & 6.14 & 1.77 & 95.27 & 9.42  \\   
		 RLR, $\alpha=$ 0.85 & 4.28 & 6.16 & 1.78 & 95.34 & 9.46  \\   
		 RLR, $\alpha=$ 0.90 & 4.23 & 6.08 & 1.76 & 95.29 & 9.33  \\   
		 RLR, $\alpha=$ 0.95 & 4.21 & 6.04 & 1.75 & 95.19 & 9.27  \\  
		 RLR, $\alpha=$ 0.98 & 4.23 & 6.08 & 1.76 & 95.24 & 9.33  \\ \hline \\  
	
	         GAM, $\alpha=$ 0.00 & 5.92 & 8.20 & 2.43 & 89.33 & 14.70  \\  
                 GAM, $\alpha=$ 0.25 & 6.16 & 8.56 & 2.53 & 89.72 & 15.33  \\  
		GAM, $\alpha=$ 0.50 & 6.59 & 9.20 & 2.72 & 91.10 & 16.54  \\
	        GAM, $\alpha=$ 0.75 & 6.78 & 9.57 & 2.80 & 93.61 & 16.88  \\ 
	        GAM, $\alpha=$ 0.85 & 6.34 & 8.99 & 2.63 & 96.42 & 15.65  \\  
	        GAM, $\alpha=$ 0.90 & 6.00 & 8.51 & 2.49 & 96.99 & 14.67  \\
	        GAM, $\alpha=$ 0.95 & 5.36 & 7.65 & 2.24 & 99.20 & 12.84  \\ 
	        GAM, $\alpha=$ 0.98 & 5.00 & 7.12 & 2.09 & 98.49 & 11.78  \\ \hline \\

	\end{tabular} 
	
	\caption{ \small Risk--reward measures of the out-of-sample daily optimal portfolio returns from 1-May-2016 to 28-April-2022.}
	\label{tab:OP_RR}
\end{table} 

\begin{table}[] 
\centering
\scalebox{0.7}{
     \begin{tabular}{ ccccccc } \hline \\ 
		
	 Model Type & Tot. Ret & Ann. Ret & AvgT  & CVaR$^{(l)}_{0.95}$ & CVaR$^{(u)}_{0.95} $&  MD  \\ \hline \\ 
	 
	EWBH                      & 136.49 & 15.67 & 0.00 & $-2.98$ & 2.51 & 33.27  \\ \hline \\  
	Factor Class 1 &  &  & & &  &   \\
	   GAM-norm-NIG-0.00  & 77.98 & 10.24 & 2.71 & $-2.34$ & 2.08 & 30.52 \\  
	   GAM-norm-NIG-0.25  & 78.47 & 10.29 & 2.71 & $-2.35$ & 2.08 & 30.28 \\ 
	   GAM-norm-NIG-0.50  & 82.09 & 10.67 & 2.71 & $-2.35$ & 2.09 & 30.21 \\
	   GAM-norm-NIG-0.75  & 83.34 & 10.79 & 2.72 & $-2.41$ & 2.10 & 30.68 \\ \hline \\  
        Factor Class 2 &  &  & & &  &   \\
           GAM-norm-NIG-0.00  & 101.95 & 12.62 & 2.72 & $-2.33$ & 2.08 & 29.64 \\ 
           GAM-norm-NIG-0.25  & 109.32 & 13.30 & 2.72 & $-2.33$ & 2.09 & 29.52  \\ 
           GAM-norm-NIG-0.50  & 119.89 & 14.25 & 2.73 & $-2.35$ & 2.11 & 29.06 \\ 
           GAM-norm-NIG-0.75  &157.87 & 17.37 & 2.76 & $-2.45$ & 2.20 & 29.22 \\ \hline \\  
        Factor Class 3 &  &  & & &  &   \\
        GAM-norm-NIG-0.00   & 108.65 & 13.24 & 2.71 & $-2.27$ & 2.07 & 28.44  \\ 
        GAM-norm-NIG-0.25   & 110.52 & 13.41 & 2.71 & $-2.26$ & 2.08 & 27.72  \\  
        GAM-norm-NIG-0.50   & 107.75 & 13.16 & 2.72 & $-2.27$ & 2.08 & 27.10  \\ 
        GAM-norm-NIG-0.75   & 134.44 & 15.50 & 2.73 & $-2.29$ & 2.14 & 25.14   \\  \hline \\  
        \end{tabular} 
        }
        \scalebox{0.8}{
        \begin{tabular}{ cccccc } \hline \\ 
		
		Model Type & IR& Sortino & STARR & Rachev & Gini  \\ \hline \\ 
		
		EWBH & 4.81 & 6.58 & 1.92 & 84.11 & 11.74  \\ \hline \\  
		
		Factor Class 1 -- Momentum&  &  &  &  &   \\
		   GAM, $\alpha=$ 0.00 & 3.98 & 5.45 & 1.63 & 88.70 & 9.56  \\ 
		   GAM, $\alpha=$ 0.25 & 3.99 & 5.46 & 1.64 & 88.52 & 9.60 \\
		   GAM, $\alpha=$ 0.50 & 4.12 & 5.62 & 1.69 & 88.56 & 9.91 \\
		   GAM, $\alpha=$ 0.75 & 4.09 & 5.54 & 1.67 & 87.25 & 9.81 \\  \hline \\  
		Factor Class 2 --  Fundamental&  &  &  &  &   \\
	           GAM, $\alpha=$ 0.00 & 4.86 & 6.69 & 2.00 & 89.31 & 11.88   \\ 
		   GAM, $\alpha=$ 0.25 & 5.11 & 7.03 & 2.10 & 89.62 & 12.50 \\
		   GAM, $\alpha=$ 0.50 & 5.43 & 7.50 & 2.23 & 89.69 & 13.32 \\
		   GAM, $\alpha=$ 0.75 & 6.30 & 8.73 & 2.57 & 89.96 & 15.53 \\  \hline \\  
		Factor Class 3 -- Technical &  &  &  &  &   \\
		   GAM, $\alpha=$ 0.00 & 5.18 & 7.20 & 2.15 & 91.13 & 12.71  \\ 
		   GAM, $\alpha=$ 0.25 & 5.24 & 7.31 & 2.18 & 91.78 & 12.91\\
		   GAM, $\alpha=$ 0.50 & 5.13 & 7.17 & 2.13 & 91.49 & 12.61 \\
		   GAM, $\alpha=$ 0.75 & 5.90 & 8.31 & 2.47 & 93.11 & 14.70 \\  \hline \\  
\end{tabular} 
        }
	\caption{ \small Comparison of summary statistics and risk--reward measures for three different GAM factor models for three choices of the reward--risk 
	trade-off parameter $\alpha$. 
	The results are based on out-of-sample daily optimal portfolio returns from 1-May-2016 to 28-April-2022.
	}
	\label{tab:OP_FK}
\end{table} 

\begin{table}[] 
	\centering
	\scalebox{0.75}{
	\begin{tabular}{ lcccccc } \hline \\ 
		
	 Model Type & Tot. Ret & Ann. Ret & AvgT  & CVaR$^{(l)}_{0.95}$ & CVaR$^{(u)}_{0.95} $&  MD  \\ \hline \\ 
	 
	 EWBH & 137.63 & 15.76 & 0.00 & $-3.01$ & $2.56$ & 35.27  \\ \hline \\  
	
         none, $\alpha=$ 0.00& 98.55 & 12.30 & 2.82 & $-2.29$ & 2.00 & 31.28  \\ 
	none, $\alpha=$ 0.25 & 97.07 & 12.16 & 2.82 & $-2.29$ & 2.00 & 31.53  \\  
	none, $\alpha=$ 0.50 & 93.77 & 11.84 & 2.82 & $-2.29$ & 2.01 & 31.14    \\   
	none, $\alpha=$ 0.75 & 88.44 & 11.31 & 2.82 & $-2.33$ & 2.05 & 31.04  \\   \hline \\
		
         RLR, $\alpha=$ 0.00 & 66.36 & 8.99 & 2.20 & $-4.16$ & 4.24 & 36.75 \\
	 RLR, $\alpha=$ 0.25 & 71.62 & 9.56 & 2.16 & $-4.22$ & 4.30 & 36.77  \\  
	 RLR, $\alpha=$ 0.50 & 79.99 & 10.45 & 2.12 & $-4.23$ & 4.32 & 36.85  \\  
	 RLR, $\alpha=$ 0.75 & 85.60 & 11.02 & 2.07 & $-4.28$ & 4.41 & 36.24   \\   \hline \\ 
	
	GAM, $\alpha=$ 0.00 & 188.82 & 19.64 & 2.86 & $-2.34$ & 2.12 & 29.91  \\ 
	GAM, $\alpha=$ 0.25 & 210.71 & 21.13 & 2.87 & $-2.40$ & 2.19 & 30.04  \\  
	GAM, $\alpha=$ 0.50 & 222.14 & 21.87 & 2.89 & $-2.46$ & 2.26 & 31.15  \\ 
	GAM, $\alpha=$ 0.75 & 240.75 & 23.04 & 2.94 & $-2.71$ & 2.46 & 34.69   \\   \hline \\ 

	\end{tabular} 
}
	\caption{ \small Summary measures of the out-of-sample daily optimal portfolio returns for the S\&P500 case from 1-May-2016 to 28-April-2022. }
	\label{tab:OP_Performance_SP500}
\end{table} 

\begin{table}[] 
\centering
\scalebox{0.85}{
	\begin{tabular}{ lcccccc } \hline \\ 
		
		Model Type                & Mean & Median & Std & Skew & ExKurt & SDev  \\ \hline \\ 
		
		 EWBH & 0.00057 & 0.00089 & $0.012$ & $-1.219$ & 22.662 & 0.009  \\ \hline \\

		none, $\alpha=$ 0.00 & 0.00046 & 0.00080 & 0.010 & $-1.344$ & 25.912 & 0.007  \\   
		none, $\alpha=$ 0.25 & 0.00045 & 0.00086 & 0.010 & $-1.343$ & 25.909 & 0.007   \\   
		none, $\alpha=$ 0.50 & 0.00044 & 0.00073 & 0.010 & $-1.314$ & 25.228 & 0.007  \\  
		none, $\alpha=$ 0.75 & 0.00042 & 0.00077 & 0.010 & $-1.288$ & 26.386 & 0.007  \\     \hline \\   
		
		 RLR, $\alpha=$ 0.00  & 0.00034 & 0.00039 & 0.018 & $-0.164$ & 5.295 & 0.013   \\ 
		 RLR, $\alpha=$ 0.25  & 0.00036 & 0.00031 & 0.018 & $-0.165$ & 5.315 & 0.013  \\   
		 RLR, $\alpha=$ 0.50  & 0.00039 & 0.00053 & 0.018 & $-0.157$ & 5.164 & 0.013  \\  
		 RLR, $\alpha=$ 0.75  & 0.00041 & 0.00054 & 0.019 & $-0.120$ & 4.844 & 0.013  \\   \hline \\ 
		
	        GAM, $\alpha=$ 0.00  & 0.00070 & 0.00108 & 0.010 & $-1.009$ & 22.999 & 0.007  \\
		GAM, $\alpha=$ 0.25  & 0.00075 & 0.00110 & 0.010 & $-0.975$ & 22.246 & 0.007  \\ 
		GAM, $\alpha=$ 0.50  & 0.00078 & 0.00113 & 0.010 & $-0.806$ & 20.460 & 0.007 \\   
		GAM, $\alpha=$ 0.75 & 0.00081 & 0.00112 & 0.011 & $-0.883$ & 18.484 & 0.008    \\ \hline \\  
		
		\end{tabular}
		} 
	\caption{ \small Main statistics of the out-of-sample daily optimal portfolio returns for the S\&P500 case from 2016-05-01 to 2022-04-28.  }
	\label{tab:OP_MainStat_SP500}
\end{table} 

\begin{table}[] 
\centering
\scalebox{0.85}{
	\begin{tabular}{ lccccc } \hline \\ 
		
 Model Type              & SR & Sortino & STARR & Rachev & Gini  \\ \hline \\ 
 EWBH                      & 4.75 & 6.46 & 1.91 & 85.02 & 11.35  \\ \hline \\  
none, $\alpha=$ 0.00 & 4.78 & 6.49 & 1.99 & 87.18 & 11.66  \\   
none, $\alpha=$ 0.25 & 4.72 & 6.41 & 1.97 & 87.44 & 11.50  \\ 
none, $\alpha=$ 0.50 & 4.60 & 6.26 & 1.92 & 88.00 & 11.14  \\
none, $\alpha=$ 0.75 & 4.32 & 5.88 & 1.80 & 88.02 & 10.44  \\ \hline \\ 
     RLR, $\alpha=$ 0.00 & 1.90 & 2.69 & 0.81 & 101.88 & 3.80  \\  
     RLR, $\alpha=$ 0.25 & 1.98 & 2.82 & 0.85 & 101.95 & 3.97  \\ 
     RLR, $\alpha=$ 0.50 & 2.14 & 3.05 & 0.92 & 102.05 & 4.29  \\ 
     RLR, $\alpha=$ 0.75 & 2.22 & 3.17 & 0.96 & 103.03 & 4.45  \\ \hline \\ 
 GAM, $\alpha=$ 0.00 & 7.23 & 10.06 & 3.00 & 90.48 & 18.57  \\ 
 GAM, $\alpha=$ 0.25 & 7.58 & 10.57 & 3.13 & 91.17 & 19.67  \\   
 GAM, $\alpha=$ 0.50 & 7.65 & 10.73 & 3.15 & 91.54 & 19.86  \\   
 GAM, $\alpha=$ 0.75 & 7.28 & 10.15 & 3.00 & 90.80 & 18.48  \\ \hline \\  		
	\end{tabular} 
}	
	\caption{ \small Risk--reward measures of the out-of-sample daily optimal portfolio returns for the S\&P500 case from 1-May-2016 to 28-April-2022.}
	\label{tab:OP_RR_SP500}
\end{table} 

\newpage

\renewcommand{\theequation}{B.\arabic{equation}}
\setcounter{equation}{0}

\section{Normal Inverse Gaussian Distributions} \label{sec:App_B}
Let $\Delta$ be a real-valued, positive semidefinite  $\left(d \times d \right)$ matrix,  $W$ a normal standard distributed $d$ dimensional random vector, $Z$ a  generalized inverse Gaussian distributed random variable  independent of $W$, and $\mu, \beta \in \mathbb{R}^{d}$ two constant vectors. Then, the $d$-dimensional random vectors $X$ follows a \textit{Multivariate Generalized Hyperbolic} (MGH) distribution if:
\begin{align}
   &X\stackrel{d}{=}\mu+Z\beta+\sqrt{Z}AW,
   \label{MGHmvm}
\end{align}
where $A \in \mathbb{R}^{d \times k}$ such that  the dispersion matrix $\Delta$ is equal to $AA'$.
In the case in which $\lambda=-\frac{1}{2}$ we obtain the multivariate normal inverse distribution $NIG_{d}\left(\alpha,\beta,\delta,\mu \right)$ with density:
{\small
\begin{align}
    & d_{NIG_{d}}\left(x; \alpha,\beta,\delta,\mu,\Delta \right)= \sqrt{\frac{2}{\pi}} \frac{\delta \alpha^{\frac{d+1}{2}}}{\left(2 \pi 
    \right)^{\frac{d}{2}}}  e^{\delta\sqrt{\alpha^{2}-\beta' \Delta \beta}}  \left(  \left(x-\mu\right)'\Delta^{-1}\left(x-\mu\right)+
    \delta^{2}\right)^{-\frac{d+1}{4}} \cdot \nonumber\\
   & \cdot K_{\frac{d+1}{2}}\left( \alpha \sqrt{ \left(x-\mu\right)'\Delta^{-1}\left(x-\mu\right)+\delta^{2} } \right)e^{\beta' \left( x-
   \mu\right)}.
\end{align}
}
MGH distributions can be parametrized in several different ways. The parametrization $\left( \lambda, \alpha,\beta, \delta,\mu,\Delta \right)$ is indeed not unique: by taking  a constant $c >0$ we have that
 $$d_{GH_{d}}\left(x; \lambda, \alpha,\beta, \delta,\mu,\Delta \right)=d_{GH_{d}}\left(x; \lambda, c^{\frac{1}{2d}}\alpha,\beta,c^{-\frac{1}{2d}} \delta,\mu,c^{\frac{1}{d}}\Delta \right).$$
This produces an identification problem in the estimation. The classical approach to overcome this issue is to restrict the dispersion matrix to have determinant equal to one: $|D|=1$.\\
Another way is to use another parametrization of the MGH distribution and to constrain the expected value of the generalized inverse gaussian distribution mixing variables $Z$ to be equal to one.  
In particular we can follow \cite{Weibel_2022} and consider  the parametrization $\left( \lambda,\bar{\alpha},\gamma,\mu,\Sigma \right)$ obtained as follows:
\begin{align}
&k=\sqrt{\frac{\delta^{2}}{\alpha^{2} -\beta' \Delta \beta }} \frac{  K_{\lambda+1}\left(\sqrt{\delta^{2} \left( \alpha^{2} -\beta' \Delta \beta \right)} \right)}{ K_{\lambda}\left(\sqrt{ \delta^{2}\left( \alpha^{2} -\beta' \Delta\right) \beta}  \right)}
\end{align}
\begin{align}
&\Sigma =k \Delta , &&\bar{\alpha} =\sqrt{\delta^{2} \left( \alpha^{2} -\beta' \Delta \beta \right)}, &&&\gamma=k\Delta \beta
\end{align}
Under this parametrization, the GIG distribution has density  
$$d_{GIG}\left(y;\lambda,\bar{\alpha} \frac{K_{\lambda}\left( \bar{\alpha}\right)}{K_{\lambda+1}\left( \bar{\alpha}\right)} ,\bar{\alpha} \frac{K_{\lambda+1}\left( \bar{\alpha}\right)}{K_{\lambda}\left( \bar{\alpha}\right)} \right),$$
 and the mean-variance normal mixture is
\begin{align}
 &X\stackrel{d}{=}\mu+Zk\Delta \beta+\sqrt{Z}AW.
 \label{MGHmvm}
\end{align}
MGH distributions are fitted to data by means of Expected Maximization (EM) algorithms. 
Consider $n$ sample data $\left[x_{1},x_{2}, \dots, x_{n} \right]$ extracted from the $d$-dimensional vector $X$. By assuming that $X$ follows an MGH distribution with parameters $\Theta=\left( \lambda,\bar{\alpha},\gamma,\mu,\Sigma \right)$, the maximum likelihood estimation is the problem of maximizing 
\begin{align}
   \ln\left[  L \left(\Theta; x_{1},x_{2}, \dots, x_{n} \right)\right]=\sum_{i=1}^{n}\ln  \left[ d_{MGH}\left(x_{i};\Theta \right) \right].    
   \label{ML}
\end{align}
If we assume we know the realizations of the latent mixing variable $Z$, we can, by means of the mean-variance Gaussian mixture representation, restated equation (\ref{ML}) as follows: 
\begin{align}
\ln\left[  \tilde{L} \left(\Theta; x_{1},x_{2}, \dots, x_{n}, z_{1}, z_{2}, \dots, z_{n}\right)\right]=&&
\sum_{i=1}^{n}\ln\left[  d_{N_{d}}\left(x_{i}|z_{i};\mu+z_{i} \gamma,z_{i} \Sigma \right)\right]  \nonumber \\
+\sum_{i=1}^{n}\ln \left[ d_{GIG}\left(z_{i};\lambda,\bar{\alpha}  \right) \right].  \label{CML}
\end{align}
As we do not observe the sample for the latent variable $Z$, an EM algorithm is implemented as a recursive two-step procedure.
A solution for the parameter vector $\Theta^{\left[1 \right]}$ is firstly computed, so that we are able to compute the conditional expectation defined by equation (\ref{CML}) given the sample $\left[x_{1},x_{2}, \dots, x_{n} \right]$ and the initial estimates of the parameters\footnote{ From a practical point of view the sample mean, the zero vector and the sample covariance matrix are used as initial guesses for $\mu$, $\gamma$ and $\Sigma$, respectively.}
  $\Theta^{\left[1 \right]}$:
\begin{align}
Q\left(\Theta;\Theta^{\left[1 \right]}\right)= \mathbb{E}\left[ \ln\left[\tilde{L} \left(\Theta; x_{1},x_{2}, \dots, x_{n}, w_{1},w_{2}, \dots, w_{n}\right)\right] |\text{ } x_{1},x_{2}, \dots, x_{n} ; \Theta^{\left[1 \right]}\right]. \label{E-step}
\end{align}
In the second step, we maximize $Q\left(\Theta;\Theta^{\left[1 \right]}\right)$ with respect to $\Theta$, so that we obtain the next set of estimated parameters 
$\Theta^{\left[2 \right]}$. The whole procedure is carried out until we observe convergence (little improvement of  $\Theta^{\left[k \right]})$. A very detailed description of the EM algorithm for MGH distributions can be found in Chapters 3.2.4 of the book of McNeil, Frey and Embrechts \cite{McNeil_2015}.

\renewcommand{\theequation}{C.\arabic{equation}}
\setcounter{equation}{0}

\section{Robust Linear Regression} \label{sec:App_C}
%
In this section, we describe the robust linear regression model. Formally, for a given date $t$ and asset $i \in \left\{ 1, \dots,I \right\} $, we define the vector  of parameters $b_{i,t}$ of size $\left( K \cdot 1\right)$ and $F_{ i , t,\tau}$ as the vector obtained by extracting the $\tau$-th row of  $F_{ i , t}$. The function $g\left( \cdot \right)$, for a particular $\tau \in \left\{ t - T_{w}  , \dots,  t -1 \right\}$, is in this case defined as follows:
\begin{align}
      & g\left( F_{i,t-1} \right) := a_{i,\tau} + F_{ i , t,\tau}' b_{i,t}.
\end{align}
Therefore, the innovation for the $i$-th asset at time $\tau$ can be written as
\begin{align}
      & h_{i,t,\tau} =  a_{i,\tau} + F_{ i , t,\tau}' b_{i,t}  +  \xi_{ i , t,\tau},
\end{align}
where $ \xi_{ i , t,\tau}$ is the corresponding residual.
Let us then define the bi-weight objective function as
\begin{equation}
    \rho(x) :=
                   \begin{cases}
                       \frac{ \kappa^{2} }{6} \left[  1 - \left( 1- \frac{x^{2}}{\kappa^{2}} \right)^{3}  \right] , \text{ for }  | x | \le \kappa, \\
                       \frac{ \kappa^{2} }{ 6 } , \text{ for } |x| > \kappa,
                   \end{cases}    
\end{equation}
and its derivative (the influence function) as
\begin{equation}
    \psi(x) :=
                   \begin{cases}
                       x \left[  1 - \frac{x^{2}}{\kappa^{2} }  \right]^{2}  , \text{ for }  | x | \le \kappa, \\
                      0 , \text{ for } |x| > \kappa,
                   \end{cases}    
\end{equation}
for some choice of the parameter $\kappa$. The parameter $\kappa$ is defined to be proportional to the scale of the response variable $h_{i,t,\tau} $, as explained in the technical documentation (see the \textit{rlm} function in \cite{Venables_2002}). The  M estimator minimises the function
\begin{align}
   \sum_{ \tau = t - T_{w} }^{t-1} \rho\left( \xi_{ i , \tau} \right) = \sum_{ \tau = t - T_{w} }^{t-1} \rho\left(  h_{i,\tau}  - F_{ i , t,\tau}' b_{i,t} \right). 
\end{align}
Differentiating the objective function and setting the derivative to 0 produces
\begin{align}
  \sum_{ \tau = t - T_{w} }^{t-1} \psi \left(  h_{i,t,\tau}  - F_{ i , t,,\tau}' b_{i,t} \right) F_{ i , t,\tau} = \mathbf{0},
\end{align}
and by defining the weight function $w_{\tau} := \frac{  \psi \left( \xi_{ i , t,\tau} \right) }{  \xi_{ i , t,\tau} }$,
the estimating equations become
\begin{align}
     \sum_{ \tau = t - T_{w} }^{t-1} w_{\tau}  \left(  h_{i,t,\tau}  - F_{ i , t,\tau}' b_{i,t} \right) F_{ i , t,\tau} = \mathbf{0}.
\end{align}
The iterative weighted least square algorithm will consist of the following steps:
\begin{enumerate}

      \item Set the iteration counter $l = 0$. 
      
               Select initial estimates $b_{i,t}^{(0)}$ and calculate the corresponding residuals, 
               $$\xi^{(0)}_{ i , t, \tau}=h_{i,t,\tau}  - F_{ i , t,,\tau}' b_{i,t}^{(0)}, $$ and the weights $w_{\tau}^{(0)} $. 
      
      Update the counter by setting $l = 1$.
      
      \item While  $ b_{i,t}^{(l)} -  b_{i,t}^{(l-1)} \approx \bf{0}$, 
               
               solve  $$\sum_{ \tau = t - T_{w} }^{t-1} w_{\tau}^{(l)}  \left(  h_{i,t,\tau}  - F_{ i , t,\tau}' b_{i,t}^{(l)} \right) F_{ i , t,\tau} = 
               \mathbf{0}$$ for $b_{i,t}^{(l)}$.
               
               Compute the residuals $\xi^{(l)}_{ i , t, \tau}=h_{i,t,\tau}  - F_{ i , t,\tau}' b_{i,t}^{(l)} $ and the weights $w_{\tau}^{(l)}$. 
               
               Update the counter by setting $l = l+1$.

\end{enumerate}

\renewcommand{\theequation}{D.\arabic{equation}}
\setcounter{equation}{0}
\section{Generalised Additive Model} \label{sec:App_D}

Let us consider again the factor model for the $i$-th asset, with $i \in \left\{ 1, \dots,I \right\} $, estimated at time $t$. 
The GAM applied in this work can be written as follows:
\begin{align}
    g( \mu_{i,t,\tau} ) =  \sum_{k=1}^{K} f_{ k }( F_{ i , t, \tau} ), 
\end{align}
where $\mu_{i,t,\tau}:= \mathbb{E}\left[ h_{i,t,\tau}\right]$. Each of the $K$ functions can be expressed as a linear combination of $q$ basis functions $u_{j}( \cdot ) $, $j=1,\dots,q$, in the following form:
\begin{align}
    f_{k}(x)=  \sum_{ j = 1}^{q} u_{j}(x) z_{j},
\end{align}
for some values of the set of parameters $z_{1}, \dots, z_{q}$. 
Here, the standardised innovation for asset $h_{i,t}$ is assumed to follow an NIG distribution, which is in the domain of attraction of the normal distribution, allowing the estimation to be obtained under Gaussian assumptions.
The basis functions have been defined as P-splines (see, for instance, \cite{Eilers_1996}) whose parameters are estimated by minimising the penalised sum of squares 
$$ \sum_{i=1}^{T_{w}} \left( h_{i,t} -   \sum_{k=1}^{K} f_{ k }( F_{ i , t, \tau} ) \right)^{2} +  \sum_{k=1}^{K} \lambda_{k} \int f_{k}^{''}(z)^{2}dz,$$
for some $\lambda_{k}>0$, $k=1,\dots,K$.
The smoothness of  each basis function is controlled by the set of parameters $\lambda_{1},\dots,\lambda_{K}$, and the second term is used to penalise models that are too `wavy' (see Ch. 3.2  in \cite{Wood_2011} for a detailed discussion). 

\end{appendices}

\bibliographystyle{acm}
\bibliography{TradingStrategy_clean}

\end{document}